\documentclass[acmtog, authorversion]{acmart}
\AtBeginDocument{%
  \providecommand\BibTeX{{%
    \normalfont B\kern-0.5em{\scshape i\kern-0.25em b}\kern-0.8em\TeX}}}

\copyrightyear{2023}
\acmYear{2023}
\setcopyright{rightsretained}
\acmConference[CHI '23]{Proceedings of the 2023 CHI Conference on Human Factors in Computing Systems}{April 23--28, 2023}{Hamburg, Germany}
\acmBooktitle{Proceedings of the 2023 CHI Conference on Human Factors in Computing Systems (CHI '23), April 23--28, 2023, Hamburg, Germany}\acmDOI{10.1145/3544548.3580695}
\acmISBN{978-1-4503-9421-5/23/04}

\usepackage{color, colortbl}
\usepackage{float}
\usepackage{multirow}
\usepackage{subfig}
\usepackage{hyperref}
\usepackage[font=small,labelfont=bf]{caption}
\begin{document}

\title[A Thematic Analysis of Dark Patterns in SNSs]{About Engaging and Governing Strategies: A Thematic Analysis of Dark Patterns in Social Networking Services}
\author{Thomas Mildner}
\orcid{0000-0002-1712-0741}
\affiliation{%
  \institution{University of Bremen}
  \city{Bremen}
  \country{Germany}
}
\email{mildner@uni-bremen.de}

\author{Gian-Luca Savino}
\orcid{0000-0002-1233-234X}
\affiliation{%
  \institution{University of St.Gallen}
  \city{St.Gallen}
  \country{Switzerland}
}
\email{gian-luca.savino@unisg.ch}

\author{Philip R. Doyle}
\orcid{0000-0002-8317-3374}
\affiliation{%
  \institution{University College Dublin}
  \city{Dublin}
  \country{Ireland}
}
\email{philip.doyle1@ucdconnect.ie}

\author{Benjamin R. Cowan}
\orcid{0000-0002-8595-8132}
\affiliation{%
  \institution{University College Dublin}
  \city{Dublin}
  \country{Ireland}
}
\email{benjamin.cowan@ucd.ie}

\author{Rainer Malaka}
\orcid{0000-0001-6463-4828}
\affiliation{%
  \institution{University of Bremen}
  \city{Bremen}
  \country{Germany}
}
\email{malaka@tzi.de}

\renewcommand{\shortauthors}{Mildner et al.}

\begin{abstract}
Research in HCI has shown a growing interest in unethical design practices across numerous domains, often referred to as ``dark patterns''. There is, however, a gap in related literature regarding social networking services (SNSs). In this context, studies emphasise a lack of users' self-determination regarding control over personal data and time spent on SNSs. We collected over 16 hours of screen recordings from Facebook's, Instagram's, TikTok's, and Twitter's mobile applications to understand how dark patterns manifest in these SNSs. For this task, we turned towards HCI experts to mitigate possible difficulties of non-expert participants in recognising dark patterns, as prior studies have noticed. Supported by the recordings, two authors of this paper conducted a thematic analysis based on previously described taxonomies, manually classifying the recorded material while delivering two key findings: We observed which instances occur in SNSs and identified two strategies --- engaging and governing --- with five dark patterns undiscovered before. 

\end{abstract}



\maketitle


\section{Introduction}
``Reading this paper makes you a better person!'' Emotionally pressuring language is just one example of a growing body of unethical and malicious design practices, often referred to as ``dark patterns''. Brignull~\cite{brignull_deceptive_nodate} first introduced the term  over a decade ago, describing dark patterns as interface design strategies that coerce or steer people into actions they would not necessarily engage in if fully informed~\cite{mathur2019}. 
Since then, significant effort has been expended on identifying, capturing, and describing examples of dark patterns, most notably in the domain of e-commerce websites where a large corpus of dark patterns has been catalogued~\cite{mathur2019}. There is, however, a need to extend this work further to capture dark patterns across other domains, such as social networking services (SNSs).
Although SNSs are used extensively across the globe, a clear view as to the types of patterns used on these sites and how these vary between major SNSs platforms has yet to be defined. Existing literature does not focus exclusively on SNSs~\cite{digeronimo2020, BongardBlanchy2021, gunawan_comparative_2021, Fritsch1141673} or focuses on specific elements only, for instance advertising dark patterns only~\cite{habib_identifying_2022} and SNSs' deletion processes~\cite{schaffner_understanding_2022}. The work does, however, emphasise an urgent need to deepen our understanding of how SNSs harm their users through unethical design practices~\cite{digeronimo2020, gunawan_comparative_2021, habib_identifying_2022, schaffner_understanding_2022}.

Gaining this understanding is a critical step in informing current efforts to legislate against such practices. Current legislation often limits regulation to data and privacy protection, in many cases neglecting potentially harmful consequences interface designs can have on individuals. For instance, both the GDPR~\cite{_gdpr_2016} and CCPA~\cite{ccpa_2018} require website providers to make their reasons for collecting data transparently whilst also offering users the option to decline any storage of personal data. However, interface designs that encourage excessive use of social media or designs that obfuscate account deletion are currently unregulated. To support future regulatory endeavours, such as a proposed draft for guidelines by the European Data Protection Board (EDPB)~\cite{edpb_guidelines_2022}, we first need to understand how SNSs use dark patterns and identify what domain-specific dark patterns might be used on these platforms.

\noindent Our work contributes to this understanding by answering two key research questions: 
\begin{itemize}
    \item[\textbf{RQ1}] What types of dark patterns are currently used in the four SNSs Facebook, Instagram, TikTok, and Twitter?
    \item[\textbf{RQ2}] Do SNSs contain dark patterns currently unique to their domain?

\end{itemize}

To address these questions, we had six HCI researchers conduct reviews across four widely used SNS platforms, based on their mobile applications: Facebook, Instagram, TikTok, and Twitter. Reviewers were assigned to identify instances of dark patterns whilst completing a range of tasks commonly carried out by users on these platforms. 
The decision fell on HCI researchers as the ability to not only recognise dark patterns but also to reflect and react to them was deemed a crucial requirement for reviewers to generate best possible results. Prior studies researching non-expert users' ability to recognise dark patterns demonstrate a significant difficulty for such tasks~\cite{BongardBlanchy2021, digeronimo2020} which we aim to mitigate by relying on experts instead.
These sessions were recorded, producing 16 hours of interaction data, which was then evaluated using thematic analysis. Based on screen recordings created during the study, instances of 44 out of 80 previously established dark patterns were identified. Thematic analysis of these identified five consistent themes the describe SNS-specific dark patterns: (1) \textit{interactive hooks}; (2) \textit{social brokering}; (3) \textit{decision uncertainty}; (4) \textit{labyrinthine navigation}; and (5) \textit{redirective conditions}. These themes were then further organised into two overarching strategies that cover more high-level incentives allowing for broader application: engaging strategies and governing strategies. These emerge from SNS-specific incentives that differ from how dark patterns are used in other domains. The strategies include two and three types of the SNS-specific dark patterns. 
Falling under the umbrella of engaging strategies are the \textit{interactive hooks} and \textit{social brokering} dark patterns, which are designed to keep users occupied and entertained with SNS for as long as possible. Whereas, SNS dark patterns that can be considered governing strategies include \textit{decision uncertainty}, \textit{labyrinthine navigation}, and \textit{redirective conditions}, which are designed to navigate users' decision-making ability on these platforms. We contribute to the current dark pattern discourse and literature by considering the impact dark patterns have on four popular SNSs, and by extending current taxonomies with instances specific to this domain. 

\section{Related Work}
In this section, we will review the relationship between research on dark patterns and work on the persuasive design in social media. We highlight the necessity to carry the dark pattern discourse over to SNSs, bridging the current gap between these two strands of research. The first two subsections present a brief overview of current dark pattern taxonomies, which were used to guide the thematic analysis conducted in this study. The overview also highlights approaches and methodologies used in previous work aimed at identifying dark patterns. Some of these are adopted in this study which builds on previous work to better understand dark patterns, specifically within the context of SNS interface design. We then continue with work that dealt with users' perception of dark patterns and their ability to recognise them in different environments. Lastly, we establish the importance of considering dark patterns in social media while following the discourse of SNS interface strategies that lead to problematic or even harmful usage behaviour. 

\subsection{Early Research On Dark Patterns}
The past decade of research into dark patterns has defined and described a comprehensive taxonomy of different types across several different domains. In recent work, Mathur et al.~\cite{Mathur2021} offer a summary of the current dark pattern landscape resulting in a dense taxonomy comprising relevant works. This taxonomy is the result of their attempt to characterise dark patterns based on the cognitive biases that they exploit. As this corpus depicts a thorough overview of past dark pattern collections, we decided to use this corpus as the basis for our study. However, we decided only to include work based on empirical research, which means that we excluded reports, such as the NCC~\cite{ncc_2018} and the CNIL~\cite{cnil_2019} that promote dark patterns to the public while recommending guidelines to make informed decisions.

One of the earliest pieces of research on dark patterns was conducted by Brignull~\cite{brignull_deceptive_nodate}, who captured and described interface strategies used to harm people through interface tricks. Producing the first taxonomy of dark pattern types, Brignull described twelve interface tricks designed to misguide users. Among these were dark patterns, such as \textit{sneak into basket}, \textit{hidden costs}, and \textit{Price Comparison Prevention}, which all operate by obscuring certain information from users of online shopping sites. The intention here is to inhibit customers' ability to make informed decisions and potentially mislead them into buying unwanted products. Alternately, the dark patterns \textit{forced continuity}, \textit{privacy zuckering}, and \textit{roach motel} use strategies that limit the options and decisions available to people when using online services. In the same year as, Brignull, Conti and Sobiesk introduced their own taxonomy 
based on findings from a twelve-month-long study aimed at cataloguing a wide range of malicious interface practices. This second taxonomy includes eleven types of dark patterns, such as \textit{coercion}, which describes interfaces that mandate users' decisions by restricting alternative options and enforcing compliance. Other techniques noted by the authors include \textit{interruptions} that interfere with a user's task flow and the \textit{obfuscation} of important information, both of which operate by hindering informed decision-making. Elsewhere, Zagal et al. examined dark patterns in video games, identifying seven types of dark patterns that specifically focus on game mechanics~\cite{zagal_dark_2013}. The research shows that while certain patterns exploit a game's ecosystem of connected users, such as \textit{social pyramid schemes} and \textit{impersonation}, others impact game-play experience like \textit{grinding} and \textit{playing by appointment}.

Elsewhere, Greenberg et al.~\cite{greenberg_dark_nodate} consider dark patterns in conjunction with proxemics theory~\cite{hall_hidden_1966}. Identifying nine types of dark patterns in total, the authors discuss interactions with potentially abusive systems in spatial environments. 
For example, the \textit{attention grabber} and \textit{disguised data collection} dark patterns could be used in the design of digital billboards and involves brands exploiting people's proximity and personal data to deliver personalised advertising to specific pedestrians as they pass by. In a similar vein – inspired by the concept of \textit {Privacy by Design} project~\cite{hustinx_privacy_2010} – Bösch et al. introduce nine further types of dark patterns, which are effectively inverse strategies to the privacy strategies developed in the Privacy by Design project. The work also highlights the role design strategies can play in manipulating users both for good and for nefarious reasons. Collectively, these early studies show that dark patterns can appear in a variety of contexts and situations, highlighting the importance of establishing a broad understanding of their origins. 

\subsection{Understanding the Origins of Dark Patterns}
Reflecting on Brignull's original work~\cite{brignull_deceptive_nodate}, Gray et al. looked to investigate how dark patterns are created in the first place. Here, researchers adopted a qualitative approach, using established taxonomies to analyse an image-based corpus of potential types of dark patterns~\cite{gray2018}. The work defines five high-level strategies that practitioners engage in when developing manipulative designs. 
For instance, the \textit{obstruction} dark pattern was used to make processes unnecessarily difficult, and incorporates Brignull's \textit{roach motel}, \textit{price comparison prevention}, and \textit{intermediate currency}~\cite{brignull_deceptive_nodate}. In later work that adopts a similar approach, Gray et al. analysed 4775 user-generated posts of the Reddit sub-forum \textit{r/assholedesign}~\cite{gray_ethical_2019}. Following multiple iterations of content analysis, the authors describe a set of six properties of ``asshole designers'' that portray malicious motivations of designers. The \textit{two-faced} property, for instance, describes designers who offer conflicting information that limits users' ability to make an informed decision. Understanding the origins of dark patterns as constraints under which practitioners work offers relevant insights into where to look when trying to recognise dark patterns anywhere.

Focusing more centrally on the frequency with which dark patterns are embedded in online interfaces, Mathur et al.~\cite{mathur2019} built a web-crawler application to collect data from over 11K shopping websites. The work identifies instances of dark patterns in more than $11\%$ of their samples. Although this percentile already presents a significant number of occurrences, the authors limited the breadth of their analysis by not analysing imagery material and suggest that many more dark patterns could be identified had other factors been taken into account. Nonetheless, guided by prior works from Gray et al.~\cite{gray2018} and Brignull~\cite{brignull_deceptive_nodate}, the authors were able to compose a taxonomy of fifteen types of dark patterns. 
Extending these works further, Mathur et al.~\cite{Mathur2021} more recently looked to identify clusters or relationships among established dark patterns, setting the basis for this work's thematic analysis. Taking the existing five high-level characteristics from their previous work ~\cite{mathur2019}, the authors added a sixth characteristic that incorporates Zagal et al.'s online gaming dark pattern \textit{disparate treatment}~\cite{zagal_dark_2013}. Their proposed model further categorises these six dark pattern characteristics into two choice architectures that distinctly affect users: (1) modification of decision space and (2) manipulation of information flow. The authors thus propose an interesting three-tiered hierarchical framework (choice architectures > high-level dark pattern characteristics > specific manifestations of dark pattern designs) under which discovered and yet-to-be-defined dark patterns can fit into. Overall, the above-mentioned work collectively describe 81 specific types of dark patterns from various domains, mostly unique in how they operate. However, we noticed an omission in regard to SNSs. Filling this gap, we build on previous work by developing a deductive codebook containing this taxonomy, which was then used during the execution of the thematic analysis.


\subsection{Recognising and Identifying Dark Patterns}
With a more central focus on end-users' perspectives of dark patterns, Di Geronimo et al.~\cite{digeronimo2020} inspected popular mobile applications sampled from the Google Play Store. The authors use a cognitive walkthrough ~\cite{nielsen_usability_1994} to identify dark patterns across a total of 240 apps, each used for ten minutes. Findings showed that $95\%$ of the tested applications contained dark patterns. In a second study, the authors evaluate users' ability to recognise dark patterns. Using an online survey, the work suggests most users had problems recognising dark patterns. Adopting a similar user-centred approach, Bongard-Blanchy et al. study peoples' awareness of manipulative interface designs and their recognition of dark patterns~\cite{BongardBlanchy2021}. 
The study, which consisted of an online survey of 413 participants, also found that 59\% of their participants were able to identify ``interface elements that can influence users' choices'' more than half of the time, suggesting that they were somewhat able to recognise dark patterns. This aligns with previous results from Di Geronimo et al.~\cite{digeronimo2020} and Maier and Harr~\cite{maier_dark_2020}. However, the work also showed that while participants understood what dark patterns were and how they might manifest, they were still deceived by them during interactions. Elsewhere, Gunawan et al.~\cite{gunawan_comparative_2021} use thematic analysis based on video recordings of online services to study differences between the various web modalities and resulting dark patterns. A contribution of their analysis is the addition of a further twelve specific types of dark patterns to the body of work. Their \textit{account deletion roadblocks} dark pattern, for instance, describes the insufficient communication between the service provider and user when the latter is trying to delete their account. Together, these works demonstrate processes for assessing and evaluating dark patterns and how they manifest. In this work, we consider a wider corpus of dark patterns on a relatively small set of applications. This allows us to offer profound insights in SNS-specific dark patterns while building on established work and their methodologies. 

\subsection{Design Strategies on SNSs}
Social media plays a major role in the daily lives of millions of people worldwide. Although prior dark pattern work has considered SNS in their research~\cite{digeronimo2020, gunawan_comparative_2021, habib_identifying_2022}, we still lack a thorough understanding of social media-specific dark patterns and the different kinds of harm they might present compared to malicious practices observed in other domains. In this context, early work by Roffarello and De Russis~\cite{monge_roffarello_towards_2022} proposes five dark patterns in Facebook and YouTube which aim to capture their users' attention. Although offering insights into SNS specific strategies, dark patterns that exploit alternative strategies than attention-capturing have yet to be described. Aside from work with an explicit focus on dark patterns, a growing body of work suggests certain uses and users of SNS may be prone to negative consequences in terms of mental health and well-being~\cite{wang2011, wang_effects_2014, beyens2020effect, shakya_association_2017}. A well-documented reason for this may be found in research comparing users' self-reported time spent on SNS to actual times, suggesting a lack of self-control and self-determination. Indeed, numerous studies have demonstrated a lack in users' ability to self-report accurately the time they spent on any SNS~\cite{junco_comparing_2013, schoenebeck_giving_2014, ernala_how_2020}. By showing that users generally spend less time on Facebook than they think while opening the application more often than realised, both Junco's and Ernala et al.'s works highlight a problem around self-awareness when it comes to frequency and length of social media use. A similar disparity was described by Mildner and Savino~\cite{mildner_ethical_2021}, where a contradiction in people's perceived and actual usage behaviour was observed among 116 participants. Most Facebook users who participated in the study admitted that they spent more time on Facebook than planned, though most also declared they had no desire to spend less time on the platform. Another interesting result of their survey is noted in participants' perceived feeling of low control over ad-related data. These findings are in line with prior research noting a general dissatisfaction in users seems to arise from limited options to protect their privacy combined with an urge to have more control over their personal data~\cite{wang_privacy_2013, dey_facebook_2012}. Investigating advertising controls on Facebook, Habib et al.~\cite{habib_identifying_2022} consider the impacts of dark patterns when conducting an online survey to support a thematic analysis identifying users' desired advertisement controls. The authors described users' difficulty in finding Facebook's Ad Preference section, to begin with. These findings affirm a prior suggestion by Gunawan et al.~\cite{gunawan_comparative_2021}, who argue that the granularity of interfaces may discourage users from making desired changes to their preferences. Both users' difficulty in self-reporting the amount of time spent on SNSs and their lack of agency to control personal settings outline unethical practices that may fall under the umbrella of dark patterns. Consequently, we see an urgent need to better understand dark patterns in SNSs considering domain-specific strategies. This necessity is further highlighted by Schaffner et al.~\cite{schaffner_understanding_2022}, who demonstrate how difficult account deletion is across 20 popular SNSs. Not only does the possibility to entirely delete an account vary depending on the modality of a particular service, but the authors further notice a difficulty among users to follow through with a deletion process. In our research, we further investigate this problem by recording and analysing reviewers' usage of four SNSs, a deletion process, allowing for detailed insights.




\section{Expert Review \& Data Collection}\label{sec:expert_review}
To collect necessary data of SNS usage for the thematic analysis, we asked six HCI experts to record their usage of four mobile SNS applications. By expanding the scope, we gain a general understanding of common practices across SNSs. The decision fell on Facebook, Instagram, TikTok, and Twitter, as they present some of the most popular SNSs while satisfying comparable user needs. To assist their review in targeting potential unethical practices, each expert reviewer was provided ten tasks that afforded them to use the mobile applications of each SNS intensively. To get a better understanding of their actions, we also asked them to narrate their decisions to retrieve data in the form of a think-aloud protocol~\cite{jaspers2004}. As we estimated that 30 minutes were required to complete all tasks, we asked each reviewer to record their usage of two of the four SNS. Thus, three independent recordings per SNS were collected. As the experiment was conducted during the COVID-19 pandemic, reviewers completed the study without supervision.

\subsection{Reviewers}

For reviewers, we reached out to HCI researchers, choosing six to investigate the four SNSs (3 female, 3 male, mean age = 28.33 years, $SD = 1.63$). All have multiple years of experience in HCI research, with backgrounds in cognitive science, computer science, and media science. At the time of the study, their years of experience varied from two to six years, with a mean of 3.83 years ($SD=1.47$).
At the time of conducting this research, all reviewers were employed as researchers at academic research faculties based in Germany, focusing on HCI related topics. Five reviewers are German citizens, while one is of Russian nationality. Except for one reviewer, participants did not have prior experience conducting dark pattern related research although all shared a general conceptualisation of the field.
Participation in this study was voluntary and without compensation. The decision fell on HCI researchers to conduct this study as regular users have been repeatedly shown to have difficulties in detecting dark patterns~\cite{digeronimo2020, BongardBlanchy2021}. In contrast, the researcher's strong expertise of usability best practices and design heuristics makes them more sensitive towards interface strategies, enabling them to uncover and discuss persuasive techniques better. Hence, their expertise allows them to identify a wider variety of dark patterns than regular users would be able to. Nonetheless, we acknowledge that this qualitative research was conducted entirely by people with strong HCI backgrounds, possibly influencing our findings and interpretations. 


\subsection{Preparation}
To counter any systematic issues caused by a particular operating system, each reviewer was provided with two smartphone devices: an iPhone X with iOS and an Android running Google Pixel 2 device. Reviewers were asked to use both devices simultaneously and to pay attention to device-specific differences when solving the tasks. Before giving them the devices, each phone was factory reset and only contained fresh installations of the SNS\footnote{Installed versions consistent throughout the study: Facebook (iOS: 321.0.0.53.119; Android: 321.0.0.37.119); Instagram: (iOS: 191.0.0.25.122; Android: 191.1.0.41.124); TikTok (iOS:19.3.0; Android: 19.3.4); Twitter (iOS: 8.69.2; Android: 8.95.0-release.00).} and some media content containing photos with creative-commons licenses. Each participant was also provided with a new phone number and email address, allowing them to create a new social media account while protecting their privacy. For the same reason, the media content was provided, as some of the tasks required the participants to create and post content. 

\subsection{Tasks}
Each evaluator was asked to execute ten tasks during their recording sessions. We decided to implement five tasks similar to Di Geronimo et al.'s~\cite{digeronimo2020} tasks as they were shown to be effective for this kind of study. To ensure coverage of SNS-specific interface problems, we also added five additional tasks tailored to this domain. Before employing the tasks to the reviewers, they were tested to ensure their accuracy and ability to cover a wide range of SNS-specific functionalities. After minor revisions following piloting, the ten tasks comprised the following exercises (whereas tasks that were taken from or worded closely to Di Geronimo et al.'s experiment are highlighted by an asterisk):

\begin{em}
\begin{enumerate}
    \item[1.] Turn on screen recording on each device.
    \item[*2.] Open the app and create an account to log in and then out.
    \item[*3.] Close and reopen the app.
    \item[4.] Create any kind of content, post it, and delete it.
    \item[5.] Follow and unfollow other accounts.
    \item[*6.] Visit the personal settings.
    \item[*7.] Visit the ad-related settings.
    \item[*8.] Use the application for its intended use (minimum of five minutes):
    \begin{enumerate}
        \item[I] Describe the natural flow of the app – what did you use it for?
        \item[II] Could you use the app as you wanted, or did some features 'guide' your interactions?
        \item[III] how easy was it to get distracted, and if so, what distracted you?
    \end{enumerate}
    \item[9.] Delete your account.
    \item[10.] Turn off screen recording and save the recording.
\end{enumerate}
\end{em}

\subsection{Procedure}
\begin{figure*}[t]
    \centering
    \includegraphics[width=0.7\textwidth]{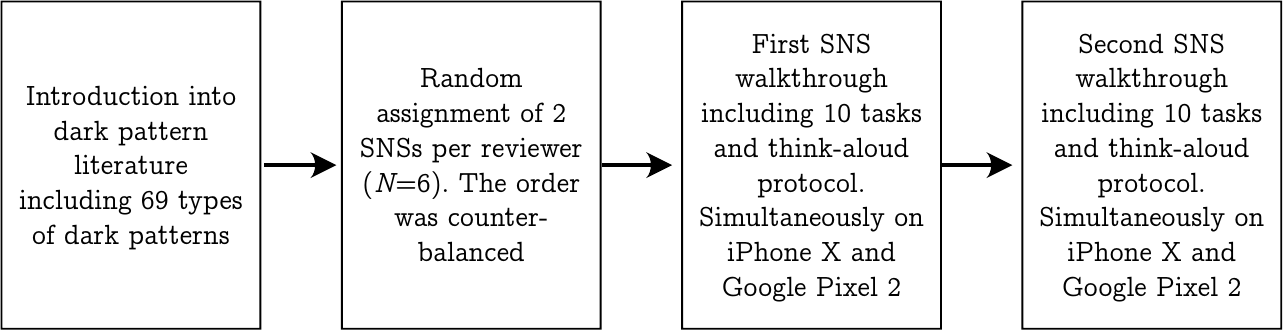}
    \caption{This flowchart describes the four steps in which our data was collected. First, reviewers received an in-depth introduction into the dark pattern literature. Each reviewer was then randomly assigned two of the four SNSs Facebook, Instagram, TikTok, and Twitter. Per SNS, the reviewers conducted a walkthrough based on 10 tasks on two devices (iOS and Android).}
    \label{fig:data-collection-flowchart}
    \Description[Flowchart of the data collection]{
    The figure visualises the process in which the six reviewers reviewed each SNS in four stages. Each stage is shown in a box with an arrow  linking it to the next. The first stage reads: "Introduction into dark pattern literature including 81 types of dark patterns."
    The second stage reads: "Random assignment of 2 SNSs per reviewer (N=6). The order was counterbalanced."
    The third stage reads: "First SNS walkthrough including 10 tasks and think-aloud protocol simultaneously on iPhone X and Google Pixel 2."
    The fourth and final stage reads: "Second SNS walkthrough including 10 tasks and think-aloud protocol. Simultaneously on iPhone X and Google Pixel 2."
    }
\end{figure*}

Before completing their walkthrough of the SNSs, reviewers were prepared for the study via a one-hour introduction session, outlining recent research on dark patterns and established taxonomies. This took the form of an online meeting. The introduction involved instructions around the 81 dark pattern types provided in the related work section, which was followed by an open discussion aimed at answering any unresolved questions reviewers may have. Afterwards, each reviewer was provided with a copy of the information, including a detailed list of dark patterns described in the introduction session. This material is included in the supplementary material of this work. Even though all participants have backgrounds in the field, this preparation ensured a common understanding of current conceptualizations of dark patterns. Information provided to reviewers also included a sheet explaining the cognitive biases particular dark patterns were designed to exploit~\cite{mathur2019}. This was included to help them identify dark patterns that were not already described in taxonomies. Each reviewer was then randomly assigned two of the four SNSs, ensuring that each SNS was examined independently three times by three different HCI researchers. The SNS applications that reviewers used were randomly assigned while ensuring that collectively reviewers did every possible sequential permutation.

Reviewers then conducted their reviews without supervision, following a detailed step-by-step guide they were provided with. The steps can be seen in Figure~\ref{fig:data-collection-flowchart}, and explain each task along with when to start and end the screen and audio recording. The manual further reminded reviewers to openly discuss their actions allowing us to retrieve data in the form of a think-aloud protocol, which we later evaluated to understand their decisions. The quality of the reviews are twofold: (1) The tasks were designed to get detailed insights into SNSs, affording reviewers to engage with the available features of each platform. (2) By recording the reviewers' commentary, we also gained information about their perception and judgement of potentially malicious artifacts. All recordings were automatically stored on each device and retrieved when reviewers returned both smartphones.


\section{Findings}
As a result of this data collection, the recordings provided over 16 hours of video and audio material, including the think-aloud commentary from reviewers. On average, each session lasted 41 minutes ($SD = 19$) for each device. Between the six participants, each SNS was thus used for an average of 247 minutes. Looking at different approaches to identifying dark patterns, we noticed varying methodologies in prior work. Of the eight taxonomies considered here, two relied on crowd-sourced or user-generated imagery material to apply qualitative methodologies, such as the constant comparative method~\cite{gray2018, gray_ethical_2019}, whereas another used a web crawler to generate a large enough corpus to apply hierarchical clustering~\cite{mathur2019}. Although highlighting important interface issues that need to be addressed, the remaining five works provide limited information to recreate their results. We decided to follow prior works and conducted a thematic analysis based on the 81 dark pattern types contained in this literature while further relying on screen recordings and audio commentary of participants. Furthermore, by limiting the scope of this research to four applications only, we were able to study each service more extensively.

\subsection{Analysis of the Data}
To answer our research questions, two researchers conducted a thematic analysis supported by the resulting data from the screen and audio recordings~\cite{braun_thematic_2019}. While the screen recordings contained the visual captures of dark patterns and interactions thereof, the audio served as complementary material, which allowed the coders to get a detailed impression of reviewers' perception and judgement of a scene. This aided the coding process in a supportive role. However, all labelling was conducted by the two coders. The thematic analysis was carried out through a combination of deductive and inductive coding~\cite{mayring_qualitative_2020} using the software ATLAS.ti~\cite{atlasti_2021}. In line with works from Di Geronimo et al.~\cite{digeronimo2020} and Gunawan et al.~\cite{gunawan_comparative_2021}, we applied codes whenever an interface was perceived problematic, rather than being driven by a concern with designer intent. This allows for the identification of dark patterns that emerge from both intentionally manipulative and unintended yet potentially harmful design choices. 
The thematic analysis was conducted by two researchers who designed and administered the experiment and aimed to identify indications of dark patterns present in the four SNSs. Although the recorded sessions were coded independently, both coders met for discussions after each session to align their interpretations of dark pattern definitions, rendering testing for inter-rater reliability unnecessary~\cite{blandford_qualitative_2016}. Additionally, all instances where coders diverged from reviewers' comments were specifically marked and later discussed among both coders. Once all sessions were analysed, they met for a final thorough discussion to establish full agreement over all 12 sessions.

\subsubsection{Creating The Deductive and Inductive Codebook}
The deductive codebook was derived from the descriptions of existing dark patterns as seen in Table \ref{tab:deductive-codes}. Each of the 81 codes included in the codebook denotes a specific type of dark pattern. That is 81 specific interface design elements that - by accident or by design - impinge on the users' autonomy. As the 12 dark patterns by Gunawan et al. were not described by the time we conducted the thematic analysis, we were not able to include them in this study. In an attempt to reduce the number of codes, we collapsed dark pattern types that shared the same name. This resulted in combining Brignull and Mathur et al.'s \textit{confirmshaming} dark patterns, which share a name and very similar descriptions, which reduced the codebook to 80 codes from an initial 81. Other candidates were also considered (\textit{privacy zuckering}~\cite{brignull_deceptive_nodate, bosch2016} and \textit{bait and switch}~\cite{brignull_deceptive_nodate, greenberg_dark_nodate}), but were not collapsed as their descriptions were deemed too distinct. We thus concluded the deductive codebook with 80 codes. 

Once the deductive codebook had been generated, a random session was chosen and then independently coded by both coders. Deductive codes were applied wherever a type of dark pattern from the taxonomy sufficed to describe a recognised problem within the interface. In those cases where an aspect of the interface was deemed a potential dark pattern but was not present within the deductive codebook, a new code was generated and added to a separate inductive codebook. Descriptions for inductive codes focused on describing the design or interaction extracted from the recordings. Once the first session was completed, the two annotating researchers met to discuss the adequacy of established inductive codes and to form a common agreement. Afterwards, overlapping codes were eventually merged to compile a single inductive codebook comprising 22 codings by using the affinity diagramming technique~\cite{blandford_qualitative_2016}.

\subsubsection{Coding Remaining Data} Both coders then proceeded to code the remaining sessions independently, following the same procedures, only interrupted to discuss and resolve inconsistencies after each session, based on Blandford et al.~\cite{blandford_qualitative_2016}. By relying on screen recordings of the interfaces, instead of sampled screenshots, we gained an important advantage that allowed us to thoroughly investigate not only dark patterns on particular frames but observe sequenced interactions that would be invisible in still images. This was further affirmed by the recordings of the think-aloud protocol. If necessary, we could stop, repeat, and compare interface situations to get a deep understanding of the interactions the reviewers performed. Because of this decision, we also noticed dark patterns that only surfaced after sequential interactions were taken. This allowed us to observe dark patterns that were not previously described and were potentially unique to SNSs.

\subsection{Deductive Codebook Analysis: Findings}

\begin{table*}[!ht]
\resizebox{0.74\textwidth}{!}{%
{\begin{tabular}{@{}lp{4cm}llllllp{4cm}llll@{}}
\cmidrule(r){1-6} \cmidrule(l){8-13}
\textbf{Author} & \textbf{Dark Pattern} & \textbf{F} & \textbf{I} & \textbf{Ti} & \textbf{Tw} &   & \textbf{Author} & \textbf{Dark Pattern} & \textbf{F} & \textbf{I} & \textbf{Ti} & \textbf{Tw} \\\cmidrule(r){1-6} \cmidrule(l){8-13} 

\multirow{14}{*}{\rotatebox[origin=c]{90}{Brignull~\cite{brignull_deceptive_nodate}}}

     &\cellcolor[HTML]{EFEFEF}Bait And Switch &\cellcolor[HTML]{EFEFEF}\huge$\bullet$&\cellcolor[HTML]{EFEFEF}\huge$\circ$&\cellcolor[HTML]{EFEFEF}\huge$\circ$&\cellcolor[HTML]{EFEFEF}\huge$\bullet$& 

                    &\multirow{9}{*}{\rotatebox[origin=c]{90} {Bösch et al.~\cite{bosch2016}}}      &\cellcolor[HTML]{EFEFEF}Address Book Leeching&\cellcolor[HTML]{EFEFEF}\huge$\bullet$&\cellcolor[HTML]{EFEFEF}\huge$\bullet$&\cellcolor[HTML]{EFEFEF}\huge$\bullet$&\cellcolor[HTML]{EFEFEF}\huge$\bullet$\\[3pt]      

     &Confirmshaming &\huge$\bullet$&\huge$\bullet$&\huge$\bullet$&\huge$\bullet$&  &               &Bad Defaults &\huge$\bullet$&\huge$\bullet$&\huge$\bullet$&\huge$\bullet$\\[3pt]
     &\cellcolor[HTML]{EFEFEF}Disguised Ads &\cellcolor[HTML]{EFEFEF}\huge$\circ$&\cellcolor[HTML]{EFEFEF}\huge$\bullet$&\cellcolor[HTML]{EFEFEF}\huge$\circ$&\cellcolor[HTML]{EFEFEF}\huge$\bullet$&  &                    &\cellcolor[HTML]{EFEFEF}Forced Registration &\cellcolor[HTML]{EFEFEF}\huge$\circ$&\cellcolor[HTML]{EFEFEF}\huge$\circ$&\cellcolor[HTML]{EFEFEF}\huge$\circ$&\cellcolor[HTML]{EFEFEF}\huge$\circ$\\[3pt]
     
     &Forced Continuity &\huge$\circ$&\huge$\circ$&\huge$\circ$&\huge$\circ$&  &                    &Hidden Legalese Stipulations &\huge$\bullet$&\huge$\bullet$&\huge$\bullet$&\huge$\bullet$\\[3pt]
     &\cellcolor[HTML]{EFEFEF}Friend Spam &\cellcolor[HTML]{EFEFEF}\huge$\circ$&\cellcolor[HTML]{EFEFEF}\huge$\circ$&\cellcolor[HTML]{EFEFEF}\huge$\circ$&\cellcolor[HTML]{EFEFEF}\huge$\circ$&  &                          &\cellcolor[HTML]{EFEFEF}Immortal Accounts &\cellcolor[HTML]{EFEFEF}\huge$\circ$&\cellcolor[HTML]{EFEFEF}\huge$\circ$&\cellcolor[HTML]{EFEFEF}\huge$\circ$&\cellcolor[HTML]{EFEFEF}\huge$\circ$\\[3pt]
     
     &Hidden Costs &\huge$\circ$&\huge$\circ$&\huge$\circ$&\huge$\circ$&  &                         &Information Milking &\huge$\bullet$&\huge$\circ$&\huge$\circ$&\huge$\circ$\\[3pt]
     &\cellcolor[HTML]{EFEFEF}Misdirection &\cellcolor[HTML]{EFEFEF}\huge$\bullet$&\cellcolor[HTML]{EFEFEF}\huge$\bullet$&\cellcolor[HTML]{EFEFEF}\huge$\bullet$&\cellcolor[HTML]{EFEFEF}\huge$\circ$&  &                   &\cellcolor[HTML]{EFEFEF}Privacy Zuckering &\cellcolor[HTML]{EFEFEF}\cellcolor[HTML]{EFEFEF}\cellcolor[HTML]{EFEFEF}\huge$\bullet$&\cellcolor[HTML]{EFEFEF}\cellcolor[HTML]{EFEFEF}\huge$\circ$&\cellcolor[HTML]{EFEFEF}\cellcolor[HTML]{EFEFEF}\huge$\bullet$&\cellcolor[HTML]{EFEFEF}\huge$\circ$\\[3pt]
     
     &Price Comparison Prevention  &\huge$\circ$&\huge$\circ$&\huge$\circ$&\huge$\circ$&  &         &Shadow User Profiles &\huge$\circ$&\huge$\circ$&\huge$\circ$&\huge$\circ$\\ \cmidrule(l){8-13}
     &\cellcolor[HTML]{EFEFEF}Privacy Zuckering &\cellcolor[HTML]{EFEFEF}\huge$\bullet$&\cellcolor[HTML]{EFEFEF}\cellcolor[HTML]{EFEFEF}\huge$\bullet$&\cellcolor[HTML]{EFEFEF}\huge$\bullet$&\cellcolor[HTML]{EFEFEF}\huge$\bullet$&  
                         
                    &\multirow{16}{*}{\rotatebox[origin=c]{90}{Gray et al.~\cite{gray2018}}}        &\cellcolor[HTML]{EFEFEF}Forced Action &\cellcolor[HTML]{EFEFEF}\huge$\bullet$&\cellcolor[HTML]{EFEFEF}\huge$\bullet$&\cellcolor[HTML]{EFEFEF}\huge$\bullet$&\cellcolor[HTML]{EFEFEF}\huge$\bullet$\\[3pt]

     &Roach Motel &\huge$\bullet$&\huge$\bullet$&\huge$\bullet$&\huge$\bullet$& &                   &\setlength\parindent{8pt}\textit{Gamification} &\huge$\bullet$&\huge$\bullet$&\huge$\bullet$&\huge$\bullet$\\[3pt]
     &\cellcolor[HTML]{EFEFEF}Sneak Into Basket &\cellcolor[HTML]{EFEFEF}\huge$\circ$&\cellcolor[HTML]{EFEFEF}\huge$\circ$&\cellcolor[HTML]{EFEFEF}\huge$\circ$&\cellcolor[HTML]{EFEFEF}\huge$\circ$& &                     &\cellcolor[HTML]{EFEFEF}\setlength\parindent{8pt}\textit{Social Pyramid} &\cellcolor[HTML]{EFEFEF}\huge$\bullet$&\cellcolor[HTML]{EFEFEF}\huge$\bullet$&\cellcolor[HTML]{EFEFEF}\huge$\bullet$&\cellcolor[HTML]{EFEFEF}\huge$\bullet$\\\cmidrule(l){9-13}
     &Trick Question &\huge$\circ$&\huge$\circ$&\huge$\circ$&\huge$\circ$& &                        &Interface Interference &\huge$\bullet$&\huge$\bullet$&\huge$\bullet$&\huge$\bullet$\\\cmidrule(r){1-6}

\multirow{14}{*}{\rotatebox[origin=c]{90}{Conti \& Sobiesk~\cite{conti_malicious_2010}}}

     &\cellcolor[HTML]{EFEFEF}Coercion  &\cellcolor[HTML]{EFEFEF}\huge$\circ$&\cellcolor[HTML]{EFEFEF}\huge$\circ$&\cellcolor[HTML]{EFEFEF}\huge$\circ$&\cellcolor[HTML]{EFEFEF}\huge$\circ$& &                             &\cellcolor[HTML]{EFEFEF}\setlength\parindent{8pt}\textit{Aesthetic Manipulation}&\cellcolor[HTML]{EFEFEF}\huge$\bullet$&\cellcolor[HTML]{EFEFEF}\huge$\bullet$&\cellcolor[HTML]{EFEFEF}\huge$\bullet$&\cellcolor[HTML]{EFEFEF}\huge$\bullet$\\[3pt]     
     
     &Confusion  &\huge$\bullet$&\huge$\circ$&\huge$\circ$&\huge$\bullet$& &                        &\setlength\parindent{8pt}\textit{False Hierarchy}&\huge$\bullet$&\huge$\bullet$&\huge$\bullet$&\huge$\bullet$\\[3pt]
     &\cellcolor[HTML]{EFEFEF}Distraction &\cellcolor[HTML]{EFEFEF}\huge$\bullet$&\cellcolor[HTML]{EFEFEF}\huge$\bullet$&\cellcolor[HTML]{EFEFEF}\huge$\bullet$&\cellcolor[HTML]{EFEFEF}\huge$\bullet$& &                   &\cellcolor[HTML]{EFEFEF}\setlength\parindent{8pt}\textit{Hidden Information} &\cellcolor[HTML]{EFEFEF}\huge$\bullet$&\cellcolor[HTML]{EFEFEF}\huge$\bullet$&\cellcolor[HTML]{EFEFEF}\huge$\bullet$&\cellcolor[HTML]{EFEFEF}\huge$\bullet$\\[3pt]
     
     &Exploiting Errors &\huge$\circ$&\huge$\circ$&\huge$\circ$&\huge$\circ$& &                     &\setlength\parindent{8pt}\textit{Preselection} &\huge$\bullet$&\huge$\bullet$&\huge$\bullet$&\huge$\bullet$\\[3pt]
     &\cellcolor[HTML]{EFEFEF}Forced Work &\cellcolor[HTML]{EFEFEF}\huge$\bullet$&\cellcolor[HTML]{EFEFEF}\huge$\bullet$&\cellcolor[HTML]{EFEFEF}\huge$\bullet$&\cellcolor[HTML]{EFEFEF}\huge$\bullet$& &                   &\cellcolor[HTML]{EFEFEF}\setlength\parindent{8pt}\textit{Toying With Emotions}&\cellcolor[HTML]{EFEFEF}\huge$\bullet$&\cellcolor[HTML]{EFEFEF}\huge$\bullet$&\cellcolor[HTML]{EFEFEF}\huge$\bullet$&\cellcolor[HTML]{EFEFEF}\huge$\bullet$\\\cmidrule(l){9-13}
     
     &Interruption &\huge$\bullet$&\huge$\bullet$&\huge$\bullet$&\huge$\bullet$& &                  &Nagging &\huge$\bullet$&\huge$\bullet$&\huge$\bullet$&\huge$\bullet$\\\cmidrule(l){9-13}
     &\cellcolor[HTML]{EFEFEF}Manipulating Navigation &\cellcolor[HTML]{EFEFEF}\huge$\bullet$&\cellcolor[HTML]{EFEFEF}\huge$\bullet$&\cellcolor[HTML]{EFEFEF}\huge$\bullet$&\cellcolor[HTML]{EFEFEF}\huge$\bullet$& &       &\cellcolor[HTML]{EFEFEF}Obstruction &\cellcolor[HTML]{EFEFEF}\huge$\bullet$&\cellcolor[HTML]{EFEFEF}\huge$\bullet$&\cellcolor[HTML]{EFEFEF}\huge$\bullet$&\cellcolor[HTML]{EFEFEF}\huge$\bullet$\\[3pt]
     
     &Obfuscation &\huge$\bullet$&\huge$\bullet$&\huge$\bullet$&\huge$\bullet$& &                   &\setlength\parindent{8pt}\textit{Intermediate Currency}&\huge$\bullet$&\huge$\bullet$&\huge$\bullet$&\huge$\bullet$\\\cmidrule(l){9-13}
     &\cellcolor[HTML]{EFEFEF}Restricting Functionalities &\cellcolor[HTML]{EFEFEF}\huge$\bullet$&\cellcolor[HTML]{EFEFEF}\huge$\bullet$&\cellcolor[HTML]{EFEFEF}\huge$\circ$&\cellcolor[HTML]{EFEFEF}\huge$\circ$& &       &\cellcolor[HTML]{EFEFEF}Sneaking &\cellcolor[HTML]{EFEFEF}\huge$\bullet$&\cellcolor[HTML]{EFEFEF}\huge$\bullet$&\cellcolor[HTML]{EFEFEF}\huge$\circ$&\cellcolor[HTML]{EFEFEF}\huge$\circ$\\[3pt]\cmidrule(l){8-13}
     
     &Shock &\huge$\circ$&\huge$\bullet$&\huge$\circ$&\huge$\circ$&

               &\multirow{8}{*}{\rotatebox[origin=c]{90}{Gray et al.~\cite{Gray2020a}}}             &Automating The User &\huge$\bullet$&\huge$\bullet$&\huge$\circ$&\huge$\bullet$\\[3pt]

     &\cellcolor[HTML]{EFEFEF}Trick&\cellcolor[HTML]{EFEFEF}\huge$\circ$&\cellcolor[HTML]{EFEFEF}\huge$\circ$&\cellcolor[HTML]{EFEFEF}\huge$\circ$&\cellcolor[HTML]{EFEFEF}\huge$\circ$& &                                  &\cellcolor[HTML]{EFEFEF}Controlling &\cellcolor[HTML]{EFEFEF}\huge$\bullet$&\cellcolor[HTML]{EFEFEF}\huge$\bullet$&\cellcolor[HTML]{EFEFEF}\huge$\bullet$&\cellcolor[HTML]{EFEFEF}\huge$\bullet$\\\cmidrule(r){1-6}

\multirow{9}{*}{\rotatebox[origin=c]{90}{Zagal et al.~\cite{zagal_dark_2013}}}

     &Grinding &\huge$\circ$&\huge$\circ$&\huge$\circ$&\huge$\circ$& &                              &Entrapping &\huge$\circ$&\huge$\circ$&\huge$\circ$&\huge$\circ$\\[3pt]
     &\cellcolor[HTML]{EFEFEF}Impersonation &\cellcolor[HTML]{EFEFEF}\huge$\circ$&\cellcolor[HTML]{EFEFEF}\huge$\circ$&\cellcolor[HTML]{EFEFEF}\huge$\circ$&\cellcolor[HTML]{EFEFEF}\huge$\circ$& &                         &\cellcolor[HTML]{EFEFEF}Misrepresenting &\cellcolor[HTML]{EFEFEF}\huge$\bullet$&\cellcolor[HTML]{EFEFEF}\huge$\bullet$&\cellcolor[HTML]{EFEFEF}\huge$\bullet$&\cellcolor[HTML]{EFEFEF}\huge$\bullet$\\[3pt]
     
     &Monetized Rivalries &\huge$\circ$&\huge$\circ$&\huge$\circ$&\huge$\circ$& &                   &Nickling-And-Diming &\huge$\circ$&\huge$\circ$&\huge$\circ$&\huge$\circ$\\[3pt]
     &\cellcolor[HTML]{EFEFEF}Pay To Skip &\cellcolor[HTML]{EFEFEF}\huge$\circ$&\cellcolor[HTML]{EFEFEF}\huge$\circ$&\cellcolor[HTML]{EFEFEF}\huge$\circ$&\cellcolor[HTML]{EFEFEF}\huge$\circ$& &                           &\cellcolor[HTML]{EFEFEF}Two Faced &\cellcolor[HTML]{EFEFEF}\huge$\circ$&\cellcolor[HTML]{EFEFEF}\huge$\circ$&\cellcolor[HTML]{EFEFEF}\huge$\circ$&\cellcolor[HTML]{EFEFEF}\huge$\circ$\\\cmidrule(l){8-13}
     
     &Playing By Appointment &\huge$\circ$&\huge$\circ$&\huge$\circ$&\huge$\circ$& 

               &\multirow{25}{*}{\rotatebox[origin=c]{90}{Mathur et al.~\cite{mathur2019}}}         &\multicolumn{5}{l}{Forced Acrtion (see Gray et al.~\cite{gray2018})}\\[3pt]

     &\cellcolor[HTML]{EFEFEF}Pre-Defined Content &\cellcolor[HTML]{EFEFEF}\huge$\circ$&\cellcolor[HTML]{EFEFEF}\huge$\circ$&\cellcolor[HTML]{EFEFEF}\huge$\circ$&\cellcolor[HTML]{EFEFEF}\huge$\circ$& &                   &\cellcolor[HTML]{EFEFEF}\setlength\parindent{8pt}\textit{Forced Enrollment}&\cellcolor[HTML]{EFEFEF}\huge$\circ$&\cellcolor[HTML]{EFEFEF}\huge$\circ$&\cellcolor[HTML]{EFEFEF}\huge$\circ$&\cellcolor[HTML]{EFEFEF}\huge$\circ$\\\cmidrule(l){9-13}
     
     &Social Pyramid Schemes &\huge$\circ$&\huge$\circ$&\huge$\circ$&\huge$\circ$& &                &Misdirection &\huge$\bullet$&\huge$\bullet$&\huge$\bullet$&\huge$\bullet$\\\cmidrule(r){1-6}

\multirow{11}{*}{\rotatebox[origin=c]{90}{Greenberg et al.~\cite{greenberg_dark_nodate}}}
     
     &\cellcolor[HTML]{EFEFEF}Attention Grabber &\cellcolor[HTML]{EFEFEF}\huge$\bullet$&\cellcolor[HTML]{EFEFEF}\huge$\bullet$&\cellcolor[HTML]{EFEFEF}\huge$\bullet$&\cellcolor[HTML]{EFEFEF}\huge$\circ$& &               &\cellcolor[HTML]{EFEFEF}\setlength\parindent{8pt}\textit{Pressured Selling}&\cellcolor[HTML]{EFEFEF}\huge$\bullet$&\cellcolor[HTML]{EFEFEF}\huge$\bullet$&\cellcolor[HTML]{EFEFEF}\huge$\bullet$&\cellcolor[HTML]{EFEFEF}\huge$\bullet$\\[3pt]
     
     &Bait And Switch&\huge$\circ$&\huge$\circ$&\huge$\circ$&\huge$\circ$& &                        &\setlength\parindent{8pt}\textit{Visual Interference}&\huge$\bullet$&\huge$\bullet$&\huge$\bullet$&\huge$\bullet$\\\cmidrule(l){9-13}
     &\cellcolor[HTML]{EFEFEF}Captive Audience&\cellcolor[HTML]{EFEFEF}\huge$\circ$&\cellcolor[HTML]{EFEFEF}\huge$\circ$&\cellcolor[HTML]{EFEFEF}\huge$\circ$&\cellcolor[HTML]{EFEFEF}\huge$\circ$& &                       &\multicolumn{5}{l}{\cellcolor[HTML]{EFEFEF}Obstruction (see Gray et al.~\cite{gray2018})}\\[3pt]
     
     &Disguised Data Collection&\huge$\circ$&\huge$\circ$&\huge$\circ$&\huge$\circ$& &              &\setlength\parindent{8pt}\textit{Hard To Cancel}&\huge$\bullet$&\huge$\bullet$&\huge$\bullet$&\huge$\bullet$\\\cmidrule(l){9-13}
     &\cellcolor[HTML]{EFEFEF}Making Personal Info. Public&\cellcolor[HTML]{EFEFEF}\huge$\circ$&\cellcolor[HTML]{EFEFEF}\huge$\circ$&\cellcolor[HTML]{EFEFEF}\huge$\circ$&\cellcolor[HTML]{EFEFEF}\huge$\circ$& &           &\cellcolor[HTML]{EFEFEF}Scarcity &\cellcolor[HTML]{EFEFEF}\huge$\bullet$&\cellcolor[HTML]{EFEFEF}\huge$\bullet$&\cellcolor[HTML]{EFEFEF}\huge$\bullet$&\cellcolor[HTML]{EFEFEF}\huge$\bullet$\\[3pt]
     
     &The Milk Factor&\huge$\circ$&\huge$\circ$&\huge$\circ$&\huge$\bullet$& &                      &\setlength\parindent{8pt}\textit{High-Demand Messages} &\huge$\circ$&\huge$\circ$&\huge$\circ$&\huge$\circ$\\[3pt]
     &\cellcolor[HTML]{EFEFEF}Unintended Relationships&\cellcolor[HTML]{EFEFEF}\huge$\circ$&\cellcolor[HTML]{EFEFEF}\huge$\circ$&\cellcolor[HTML]{EFEFEF}\huge$\circ$&\cellcolor[HTML]{EFEFEF}\huge$\circ$& &               &\cellcolor[HTML]{EFEFEF}\setlength\parindent{8pt}\textit{Low-Stock Messages} &\cellcolor[HTML]{EFEFEF}\huge$\circ$&\cellcolor[HTML]{EFEFEF}\huge$\circ$&\cellcolor[HTML]{EFEFEF}\huge$\circ$&\cellcolor[HTML]{EFEFEF}\huge$\circ$\\\cmidrule(l){9-13}
     
     &We Never Forget&\huge$\circ$&\huge$\circ$&\huge$\circ$&\huge$\circ$& &                        &\multicolumn{5}{l}{Sneaking (see Gray et al.~\cite{gray2018})}\\\cmidrule(r){1-6}
\textbf{Legend:}&\textbf{F} - Facebook& & & & & &                                                   &\cellcolor[HTML]{EFEFEF}\setlength\parindent{8pt}\textit{Hidden Subscriptions} &\cellcolor[HTML]{EFEFEF}\huge$\circ$&\cellcolor[HTML]{EFEFEF}\huge$\circ$&\cellcolor[HTML]{EFEFEF}\huge$\circ$&\cellcolor[HTML]{EFEFEF}\huge$\circ$\\\cmidrule(l){9-13}
     
     &\textbf{I} - Instagram & & & & & &                                                            &Social Proof &\huge$\bullet$&\huge$\bullet$&\huge$\bullet$&\huge$\bullet$\\[3pt]
     &\textbf{Ti} - TikTok & & & & & &                                                              &\cellcolor[HTML]{EFEFEF}\setlength\parindent{8pt}\textit{Activity Notifications} &\cellcolor[HTML]{EFEFEF}\huge$\circ$&\cellcolor[HTML]{EFEFEF}\huge$\circ$&\cellcolor[HTML]{EFEFEF}\huge$\circ$&\cellcolor[HTML]{EFEFEF}\huge$\circ$\\[3pt]
     
     &\textbf{Tw} - Twitter & & & & & &                                                             &\setlength\parindent{8pt}\textit{Testimonials} &\huge$\circ$&\huge$\circ$&\huge$\circ$&\huge$\circ$\\\cmidrule(l){9-13}
     &&&&&&&                                                                                        &\cellcolor[HTML]{EFEFEF}Urgency &\cellcolor[HTML]{EFEFEF}\huge$\bullet$&\cellcolor[HTML]{EFEFEF}\huge$\bullet$&\cellcolor[HTML]{EFEFEF}\huge$\bullet$&\cellcolor[HTML]{EFEFEF}\huge$\bullet$\\[3pt]
     
     &&&&&&&                                                                                        &\setlength\parindent{8pt}\textit{Countdown Timer} &\huge$\circ$&\huge$\circ$&\huge$\circ$&\huge$\circ$\\[3pt]
     &&&&&&&                                                                                        &\cellcolor[HTML]{EFEFEF}\setlength\parindent{8pt}\textit{Limited-Time Messages} &\cellcolor[HTML]{EFEFEF}\huge$\circ$&\cellcolor[HTML]{EFEFEF}\huge$\circ$&\cellcolor[HTML]{EFEFEF}\huge$\circ$&\cellcolor[HTML]{EFEFEF}\huge$\circ$\\[3pt]

\\ \bottomrule

\end{tabular}
    }
}
\caption{This table offers an overview of all 80 deductive codes used in the thematic analysis, sorted by authors. We only considered original sources for dark patterns to rely on unique codes and avoiding redundancy. As both Gray et al.~\cite{gray2018} and Mathur et al.~\cite{mathur2019} include both high-level and low-level definitions, we used indentation to highlight categorical differences for these types of dark patterns. For transparency, we added three of Mathur et al.'s~\cite{mathur2019} high-level dark pattern categories, which were carried over from Gray et al.~\cite{gray2018} and labeled accordingly. The codes were applied in four SNSs (\textbf{F} - Facebook, \textbf{I} - Instagram, \textbf{Ti} - TikTok, and \textbf{Tw} - Twitter). Codes that were identified by either of the two coders conducting the thematic analysis are indicated with ``{\huge$\bullet$}'' whereas ``{\huge$\circ$}'' indicates that a code for a dark pattern was not found in the SNSs.}
\label{tab:deductive-codes}
\end{table*}

Answering our first research question, we applied 44 out of the 80 deductive codes across our dataset, highlighting how many distinct dark patterns occurred across the four different SNSs. Table~\ref{tab:deductive-codes} displays all dark patterns and shows for which SNS they were applied during the analysis. Of these 44 that were applied, we observed 32 instances on each of the four SNSs. Regarding the original scope of individual dark pattern taxonomies, we noticed that
the dark patterns described by Conti and Sobiesk~\cite{conti_malicious_2010} and Gray et al.~\cite{gray2018} were noticed most commonly in the context of SNS, suggesting easier applicability of these taxonomies. Interestingly, these sets of dark patterns were both originally created by analysing a wide range of interfaces rather than focusing on a specific domain. Further, those described by Gray et al.~\cite{gray2018} even subsumed dark pattern types formerly described by Brignull~\cite{brignull_deceptive_nodate}, generalising them even further. 
We also investigated which SNSs feature the largest variety of dark pattern types. Facebook, which contained 41 different types of dark patterns, exhibited the most variety, followed by Instagram, featuring 39, Twitter, where 35 different types of dark patterns were observed, and finally Tiktok, with 37 different types identified.

While this result highlights that SNS seem to make use of a wide variety of dark pattern types, 36 dark pattern codes were still left unused, implying that these were not appropriate to describe dark patterns in SNSs. Three groups of dark patterns from the initial 80 were not or only rarely applied: the proxemic dark patterns by Greenberg et al.~\cite{greenberg_dark_nodate}, the gaming dark patterns by Zagal et al.~\cite{zagal_dark_2013}, and the e-commerce dark patterns by Mathur et al.~\cite{mathur2019}. 
Many of these dark patterns were generated by describing design decisions in a specific domain or context, making it almost impossible to apply them elsewhere. One example would be Greenberg's \textit{captive audience} dark pattern, which requires a "person [to] [enter] a particular area"~\cite{greenberg_dark_nodate} relying heavily on the actual proximity, hence why they are less applicable to a domain as ubiquitous as SNSs. 


Beyond overly-specific dark patterns, others share a name, albeit with varying definitions. As a result, two dark patterns with the same name do not necessarily apply identically to SNS. For example, \textit{privacy zuckering} was described by Brignull~\cite{brignull_deceptive_nodate} and Bösch et al.~\cite{bosch2016}. While Bösch et al. focuses specifically on sharing more data than a user intends to give based on privacy settings, Brignull does not make this specification and thus leaves more room for interpretation and application. Hence, Brignull's version of the dark pattern was applied across all four SNS during our analysis, while the version by Bösch et al. was only applied explicitly on Facebook and TikTok. In other cases, some names of dark pattern promise applicability in SNSs but were actually never applied as their definition hindered accurate coding. An example is illustrated by the dark pattern \textit{immortal accounts} by Bösch et al.~\cite{bosch2016}. While the name of the dark pattern suggests that user accounts are (almost) impossible to delete, its definition refers to service providers requiring new users to sign up for accounts to use their service.

\subsection{Inductive Codebook Analysis: Findings}
Although 44 codes from the deductive codebook could be applied in the SNSs, often, new codes were required to describe interface issues not yet covered. We decided early on to stick rigidly to established definitions of the types of dark patterns when applying deductive codes to avoid ambiguity among coders. During the coding of a first sample, we noticed issues were close in nature to certain deductive codes, but narrow wording hindered precise usage in different contexts. For instance, this was the case with the \textit{confirmshaming} code. The underlying dark pattern, first coined by Brignull~\cite{brignull_deceptive_nodate}, describes texts that guilt users into certain actions based on shameful language. Applying an inverse strategy, we noticed various cases in which language was used in the form of positive encouragement to steer users towards a certain direction. We, therefore, created an additional code, \textit{persuasive langugage}, that we defined bidirectionally describing any instance where language is used to push decisions. In total, we defined 22 unique codes describing problematic interactions or otherwise questionable interfaces that could not be coded using deductive codes (the entire inductive codebook, including code descriptions, are provided in Appendix ~\ref{app:inductive-codebook}). As seen in Table~\ref{tab:inductive-codes}, the 22 inductive codes did not occur across each of the four SNSs equally. Yet, 16 codes were applied to recording sessions on Facebook, 15 to TikTok and Twitter, while 14 were applied on Instagram sessions, showing a similar propagation of dark patterns across SNSs. Once all twelve sessions were analysed, we applied axial coding to the 22 codes. This resulted in the identification of five themes that describe the various types of dark patterns that are specific to SNSs. Answering our second research question, these themes encompass: (1) \textit{interactive hooks}; (2) \textit{social brokering}; (3) \textit{decision uncertainty}; (4) \textit{labyrinthine navigation}; and (5) \textit{redirective conditions}. 
Moreover, we were able to assign these five themes to two broader strategies, to describe practitioners' and SNSs' intentions to navigate users' decision-making: (1) Engaging strategies and (2) Governing strategies. Engaging strategies envelope the themes \textit{interactive hoooks} and \textit{social-broking}, whilst governing strategies incorporate \textit{decision uncertainty}, \textit{labyrinthine navigation}, and \textit{redirective conditions}. Table~\ref{tab:inductive-codes} provides a complete summary of these findings, including the 22 inductive codes, resulting in five themes and overarching strategies. 


\begin{table*}[!ht]
\begin{tabular}{llrlcccc}
\hline
\multicolumn{1}{c}{\textbf{Strategy}} & \multicolumn{1}{c}{\textbf{Theme}} & \multicolumn{1}{c}{\textbf{No.}} & \multicolumn{1}{c}{\textbf{Inductive Code}} & \textbf{F} & \textbf{I} & \textbf{Ti} & \textbf{Tw}\\ \hline
& \multicolumn{1}{l}{} & 1. & \multicolumn{1}{l}{Addictive Design} & {\huge$\circ$} & {\huge$\bullet$} & {\huge$\bullet$} & {\huge$\bullet$}\\
& \multicolumn{1}{l}{} & 2. & \multicolumn{1}{l}{Autoplay Content} & {\huge$\bullet$} & {\huge$\bullet$} & {\huge$\bullet$} & {\huge$\bullet$} \\
& \multicolumn{1}{l}{} & 3. & \multicolumn{1}{l}{Fear Of Missing Out} & {\huge$\circ$} & {\huge$\circ$} & {\huge$\circ$} & {\huge$\bullet$}\\
& \multicolumn{1}{l}{} & 4. & \multicolumn{1}{l}{Gamification} & {\huge$\bullet$} & {\huge$\bullet$} & {\huge$\bullet$} & {\huge$\bullet$} \\
& \multicolumn{1}{l}{} & 5. & \multicolumn{1}{l}{Infinite Scrolling} & {\huge$\bullet$} & {\huge$\bullet$} & {\huge$\bullet$} & {\huge$\bullet$} \\
& \multicolumn{1}{l}{} & 6. & \multicolumn{1}{l}{Pull To Refresh} & {\huge$\bullet$} & {\huge$\bullet$} & {\huge$\bullet$} & {\huge$\bullet$} \\
& \multicolumn{1}{l}{\multirow{-7}{*}{Interactive Hook}} & 7. & \multicolumn{1}{l}{Reduced Friction} & {\huge$\bullet$} & {\huge$\bullet$} & {\huge$\bullet$} & {\huge$\bullet$} \\ \cline{2-8} 

& \multicolumn{1}{l}{} & 8. & \multicolumn{1}{l}{False Content Customisation} & {\huge$\bullet$} & {\huge$\circ$} & {\huge$\circ$} & {\huge$\circ$} \\
& \multicolumn{1}{l}{} & 9. & \multicolumn{1}{l}{Regression Toward The Mean} & {\huge$\bullet$} & {\huge$\circ$} & {\huge$\bullet$} & {\huge$\bullet$} \\
\parbox[t]{2mm}{\multirow{-10}{*}{\rotatebox[origin=c]{90}{\textbf{Engaging Strategies}}}} & \multicolumn{1}{l}{\multirow{-3}{*}{Social Brokering}} & 10. & \multicolumn{1}{l}{Social Connector} & {\huge$\bullet$} & {\huge$\bullet$} & {\huge$\bullet$} & {\huge$\circ$} \\ \hline

& \multicolumn{1}{l}{} & 11. & \multicolumn{1}{l}{Decision Uncertainty} & {\huge$\bullet$} & {\huge$\circ$} & {\huge$\circ$} & {\huge$\bullet$} \\
& \multicolumn{1}{l}{} & 12. & \multicolumn{1}{l}{Clinging To Accounts} & {\huge$\bullet$} & {\huge$\bullet$} & {\huge$\bullet$} & {\huge$\bullet$} \\
& \multicolumn{1}{l}{\multirow{-3}{*}{Decision Uncertainty}} & 13. & \multicolumn{1}{l}{Persuasive Language} & {\huge$\bullet$} & {\huge$\bullet$} & {\huge$\bullet$} & {\huge$\bullet$} \\ \cline{2-8}

& \multicolumn{1}{l}{} & 14. & \multicolumn{1}{l}{External Solution Search} & {\huge$\circ$} & {\huge$\bullet$} & {\huge$\circ$} & {\huge$\circ$} \\
& \multicolumn{1}{l}{} & 15. & \multicolumn{1}{l}{Labyrinth} & {\huge$\bullet$} & {\huge$\bullet$} & {\huge$\circ$} & {\huge$\bullet$} \\
& \multicolumn{1}{l}{\multirow{-3}{*}{Labyrinthine Navigation}} & 16. & \multicolumn{1}{l}{Hidden In Plain Sight}  & {\huge$\bullet$} & {\huge$\bullet$} & {\huge$\bullet$} & {\huge$\bullet$}\\ \cline{2-8}

& \multicolumn{1}{l}{} & 17. & \multicolumn{1}{l}{Auto Accept Third Party Terms} & {\huge$\circ$} & {\huge$\circ$} & {\huge$\circ$} & {\huge$\bullet$} \\
& \multicolumn{1}{l}{} & 18. & \multicolumn{1}{l}{Decision Governing} & {\huge$\bullet$} & {\huge$\bullet$} & {\huge$\circ$} & {\huge$\circ$} \\
& \multicolumn{1}{l}{} & 19. & \multicolumn{1}{l}{Forced Access Granting} &  {\huge$\circ$} & {\huge$\circ$} & {\huge$\bullet$} & {\huge$\circ$} \\
& \multicolumn{1}{l}{} & 20. & \multicolumn{1}{l}{Forced Dialogue Interaction} & {\huge$\bullet$} & {\huge$\circ$} & {\huge$\bullet$} & {\huge$\circ$} \\
\parbox[t]{2mm}{\multirow{-10}{*}{\rotatebox[origin=c]{90}{\textbf{Governing Strategies}}}} & \multicolumn{1}{l}{\multirow{-4}{*}{Redirective Condition}} & 21. & \multicolumn{1}{l}{Forced Grace Period} & {\huge$\bullet$} & {\huge$\bullet$} & {\huge$\bullet$} & {\huge$\bullet$} \\ 

& \multicolumn{1}{l}{} & 22. & \multicolumn{1}{l}{Plain Evil} & {\huge$\circ$} & {\huge$\circ$} & {\huge$\bullet$} & {\huge$\circ$} \\ \hline
 
& & \multicolumn{1}{l}{} & \multicolumn{1}{r}{\textbf{Total}} & 16 & 14 & 15 & 15 \\ \cline{3-8}
\end{tabular}
\caption{This table lists inductive codes and their presence within the four SNSs Facebook (\textbf{F}), Instagram (\textbf{I}), TikTok (\textbf{Ti}), and Twitter (\textbf{Tw}). Inductive codes observed by either of the two coders in a particular SNS session are visualised with ``{\huge$\bullet$}'' while ``{\huge$\circ$}'' implies that a code was not found. This table further shows the themes each inductive code was assigned to based on axial coding of the data and the high-level strategies they subscribe to - engaging strategies and governing strategies. To align with prior research of this field, the developed themes are later referred to as types of dark patterns.}
\label{tab:inductive-codes}
\end{table*}




\subsection{Engaging Strategies}
In the context of SNSs, engaging strategies cover dark patterns where the goal is to keep users occupied and entertained for as long as possible. Prior dark patterns described by Roffarello and Russis~\cite{monge_roffarello_towards_2022} fall under this strategy alongside Lukoff et al.'s~\cite{lukoff_what_2018} discussion of potentially existing \textit{attention-capture} dark patterns, which the OECD recently integrated~\cite{oecd__2022_dark}.
As Table~\ref{tab:inductive-codes} indicates, the overall count of engaging strategies and subordinated codes is higher compared to those assigned to governing strategies, which is in line with research on people's motivation to use SNSs~\cite{wong_motivations_2017}. Emerging from the five themes, we identified two SNS-specific dark patterns - \textit{interactive hooks} and \textit{social brokering} - that subscribe to the engaging strategies (see Table~\ref{tab:inductive-codes}).

\subsubsection{Interactive Hooks}
\begin{figure*}[t]
    \centering
    \subfloat[\centering Screenshot from Facebook]
    {{\includegraphics[width=0.24\textwidth]{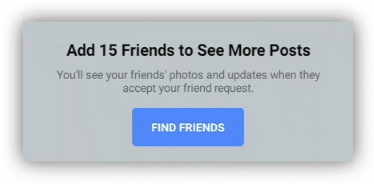}}}\hfill%
    \subfloat[\centering Screenshot from Instagram]
    {{\includegraphics[width=0.24\textwidth]{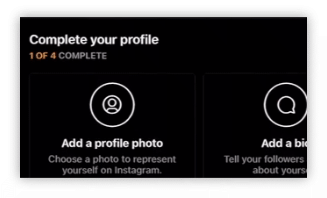}}}\hfill%
    \subfloat[\centering Screenshot from TikTok]
    {{\includegraphics[width=0.24\textwidth]{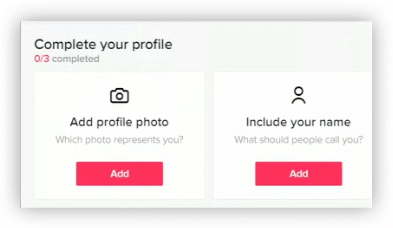}\label{fig:gamification-tiktok}}}\hfill%
    \subfloat[\centering Screenshot from Twitter]
    {{\includegraphics[width=0.24\textwidth]{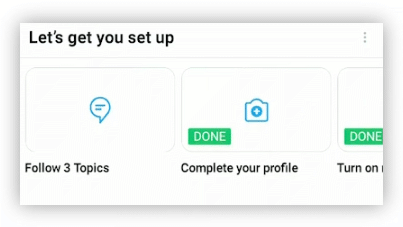}\label{fig:gamification-twitter}}}\hfill%
    \caption{Four examples of \textit{interactive hooks}. Each sub-figure contains gamification elements that galvanise users to widen their social networks or publish more information about themselves.}
    \Description[Screenshots of the four mobile apps with interactive hook dark patterns]{
    This figure contains four sub-figures of four examples of the interactive hooks dark pattern. Figure 2a) shows a screenshot from Facebook asking a user to add 15 friends to see more posts. A big blue button below states find friends nudging users to connect to more people.
    Figure 2b) shows an interface from Instagram asking users to complete their profiles. A gamification-like element indicates progress currently at 1 of 4 complete.  Below are options with icons to add a photo or add biographic data.
    Figure 2c) shows a picture from TikTok. Similar to the Instagram interface, it asks users to complete their profile with a similar progress bar showing 0 of 3. Below are options to add a profile photo or include a name as an addition to the user name. Buttons for both options are bright red.
    Figure 2d) stems from Twitter. The interface announces: let's get you set up. Below the statement are boxes with game-like tasks. The first requires the user to follow three topics. The second asks them to complete their profile. A green 'Done' text over the second suggests that this task has been accomplished.}
    \label{fig:gamification-screenshots}
\end{figure*}

We define \textit{interactive hooks} as design mechanisms that use rewarding schemes to keep users entertained and spend more time on a service. Throughout our recordings, we found multiple cases where such mechanisms were implemented. For example, some form of \textit{gamification} elements~\cite{simoes_social_2013} were coded within each SNS, galvanising users to share more information about themselves or connect to new people (see Figure~\ref{fig:gamification-screenshots}). In another example, we found that many artefacts utilise \textit{addictive mechanisms}~\cite{montag_addictive_2019} that often provide seemingly infinite content, further coded with \textit{pull-to-refresh}, \textit{infinite scrolling}, and \textit{auto playing} media. With this finding, we confirm previous results of Lukoff et al.~\cite{lukoff_how_2021} who looked at auto-playing mechanisms on YouTube. We extend their work by also considering four alternative SNSs. By relying on a sequence of interactions, these addictive mechanisms can only be described with difficulty from still images, demonstrating the complexity in which dark patterns manifest in SNSs.

\subsubsection{Social Brokering}
\begin{figure}[t]
    \centering
    \includegraphics[width=0.45\textwidth]{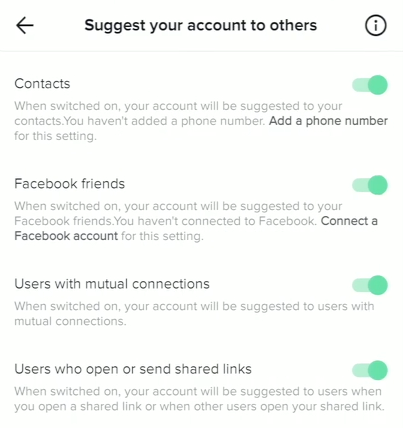}
    \caption{Example of \textit{social brokering} from TikTok where settings are pre-set to suggest an account within and outside the SNS. The interface includes \textit{bad defaults} and \textit{privacy zuckering} dark patterns}
    \label{fig:social-connector-tiktok}
    \Description[First example of a social brokering dark pattern from TikTok]{The image shows an interface from TikTok. At the top, it reads: "Suggest your account to others". Four options are seen: First contacts: meaning that one's account is suggested to others if a phone number is provided. Second are "Facebook friends", allowing TikTok to connect to a Facebook account. The third option reads users with mutual connections, and the fourth is users who open or send shared links. Further deploying the bad defaults and privacy zuckering dark patterns, all options are active.}
\end{figure}

\begin{figure}[t]
    \centering
    \includegraphics[width=0.45\textwidth]{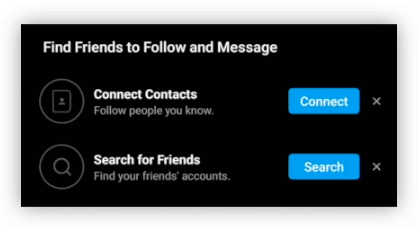}
    \caption{Example of \textit{social brokering} from TikTok where settings are pre-set to suggest an account within and outside the SNS.}
    \label{fig:social-connector-instagram}
    \Description[Second example of a social brokering dark pattern from TikTok]{This interface from TikTok asks users to find friends to follow and message them. Two options are provided: a connect to contacts option and a search for friends option. Both have a big blue button to follow the suggestion, whereas light grey X-buttons would allow dismissal.}
\end{figure}

\begin{figure}[t]
    \centering
    \includegraphics[width=0.45\textwidth]{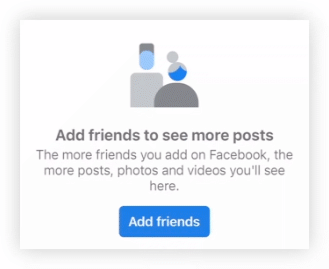}
    \caption{Example of \textit{social brokering} from Facebook. Once users reach the bottom of their timeline content, this message appears, nudging them to add friends to see new posts.}
    \label{fig:social-connector-facebook}
    \Description[Third example for social brokering dark pattern from Facebook]{This Facebook screenshot asks users to add friends to see more posts. In a light grey text, it uses persuasive language to nudge them by saying that more friends mean more photos and videos to be seen. A big blue button to add friends is provided.}
\end{figure}

We define \textit{social brokering} as design mechanisms that nudge users to create multiple connections with people (e.g. based on similar characteristics) while suggesting new people to connect to, leading users to share more than they may want to a wider public. The name is inspired by agents whose aim is to facilitate connections between potential (transaction-) partners - here in the context of social networks. Although gamification strategies of the \textit{interactive hooks} pattern already encourage users to increase their social connections, we found a range of artefacts specifically designed for this purpose codes as \textit{social connector}. Moreover, we noticed that each SNS customised content presented based on the reviewers' usage behaviour. While in some instances the news feed content appeared a poor fit for the reviewers' preferences (coded as \textit{false content customisation}), we also found very popular content reappearing across the recordings, captured in the code \textit{regression toward the mean}. Visualised in Figure~\ref{fig:social-connector-tiktok}, Brignull's bad defaults pattern~\cite{brignull_deceptive_nodate} is exploited to promote one's account to other users also outside the platform itself. Figure~\ref{fig:social-connector-instagram} shows how Instagram nudges its users to upload their device's local contacts to connect to more accounts quickly. However, by giving away details from private contacts, SNS providers can also store information from people who do not use their services without getting their consent first. In a similar case, Facebook uses persuasive language by telling users to add friends in order to see more content (see Figure~\ref{fig:social-connector-facebook}).

\subsection{Governing Strategies}
Governing strategies describe interface designs that navigate users' decision-making towards the designers' and/or platform providers' goals. Essentially, these are designed to control or govern user behaviour. While existing dark patterns, such as \textit{interface interference}~\cite{gray_ethical_2019}, fit this strategy's scope, we shine a light on not yet discussed dark pattern types: (1) \textit{decision uncertainty}; (2) \textit{labyrinthine navigation}; (3) and \textit{redirective conditions}. All these strategies share a limitation of users' agency as SNS providers override users' goals with their own incentives.

\subsubsection{Decision Uncertainty}
\begin{figure}[t]
    \centering
    \includegraphics[width=0.45\textwidth]{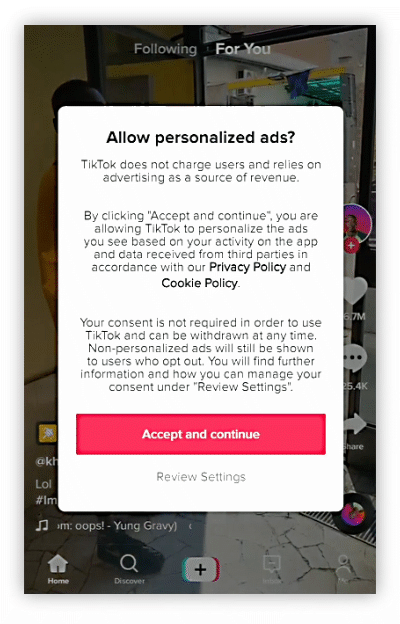}
    \caption{Example of \textit{decision uncertainty} from TikTok. After first logging in, users of the SNS are asked to choose preferences for their ad-related settings. Meanwhile, video and audio media are playing in the background, complicating the interaction. The interface further contains \textit{interface interference, visual interference} and hidden-legalese stipulation dark patterns.}
    \label{fig:decision-uncertainty-tiktok}
    \Description[Example of a decision uncertainty dark pattern from TikTok]{This TikTok interface shows an interface prompt blocking further interaction. It asks users to allow personalised ads. Three paragraphs of light grey text yield links to legal information whereas a big red button reads: "Accept and continue" while a "review settings" button is obfuscated by a light grey font colour. Not visible in this screenshot is the situation where this prompt appeared during the first initialisation of the application while loud audio and video played in the background, overwhelming users.}
\end{figure}

We define \textit{decision uncertainty} as strategies that are confusing to users by diminishing their ability to assess situations, leaving the user clueless as to what is expected of them or what options are available. Most similar to this dark pattern is the \textit{confusing} dark pattern by Conti and Sobiesk~\cite{conti_malicious_2010}. However, in SNS, the strategy is not necessarily limited to incomprehensible questions or information but includes other elements, like \textit{distraction}~\cite{conti_malicious_2010}, that overwhelm users. Interface elements of the \textit{decision uncertainty} code were so striking that we decided to promote it to a theme. Other elements obscured decision-making further. Some interfaces obfuscated the account deletion process (coded with \textit{clinging to accounts}) while others used \textit{persuasive language} to confuse users, similar to the \textit{confirmshaming} dark pattern~\cite{brignull_deceptive_nodate}. We found a quite unique design choice TikTok users faced when first logging on to the platform: When first opening the app after logging in, users are prompted with an interface asking them to choose preferred ad-related settings. While making their decision, both video and audio media is running in the background, contributing to cognitive overload. During the recordings, we noticed that reviewers of the data collection quickly complied with the platform's preference that utilises interface interference~\cite{gray_ethical_2019} and visual interference~\cite{mathur2019} dark patterns (see Figure~\ref{fig:decision-uncertainty-tiktok}). In a second example, Twitter users trying to delete their accounts will only find an option to deactivate it. Although it is possible to fully delete their account by following this path, the wording obfuscates this possibility.

\subsubsection{Labyrinthine Navigation}
We define \textit{labyrinthine navigation} as nested interfaces that are easy to get lost in, disabling users from choosing preferred settings. This pattern is often seen in SNS settings menus. Related to \textit{manipulating navigation}~\cite{conti_malicious_2010}, but not necessarily steering users towards a designer's objective. Instead, this dark pattern describes interface architectures, such as menus, that users will easily get lost in, leaving them unsuccessful in achieving their goals. While recording the data, especially tasks six (visit the personal settings) and seven (visit the ad-related settings), surfaced difficulties for reviewers to find specific settings, some of them coded \textit{hidden in plain sight} camouflaged between a wide collection of other options. We noticed this issue across all four SNSs. In a particular case, we noticed one participant using Instagram to switch applications to an online search engine to look up the solution for how to find a specific setting. 

\subsubsection{Redirective Conditions}
\begin{figure}[t]
    \centering
    \includegraphics[width=0.45\textwidth]{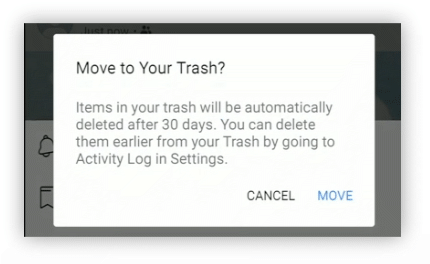}
    \caption{Example of \textit{redirective conditions} from Facebook. When trying to delete a post, Facebook will instead keep the item stored for another 30 days.}
    \label{fig:redirective-conditions-facebook}
    \Description[First example of a redirective conditions dark pattern from Facebook]{This interface from Facebook wants a user to reaffirm whether they want to move an item to the trash. In grey text it reads that items will be stored for 30 days before they will be deleted, denying the user immediate deletion.}
\end{figure}

\begin{figure}[t]
    \centering
    \includegraphics[width=0.45\textwidth]{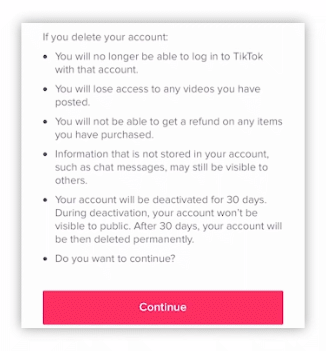}
    \caption{Example of \textit{redirective conditions} from TikTok. Accounts will be kept and able to be restored for 30 days without giving users the choice to bypass this decision.}
    \label{fig:redirective-conditions-tiktok}
    \Description[Second example of a redirective conditions dark pattern from TikTok.]{This TikTok interface stems from an account deletion process. It shows a list of bullet points of reasons to stay, persuading the user to cancel this action. At the bottom, a big red continue button is placed.}
\end{figure}

We define \textit{redirective condition} as choice limitations that force users to overcome unnecessary obstacles before being able to achieve their goals. Redirective conditions usually favour the objectives of the SNS. The \textit{forced action} dark pattern~\cite{gray_ethical_2019} plays an important role here, as users are required to first comply with certain demands before being able to do what they want. In the context of SNS, this dark pattern includes passive functionalities, like the restriction of services that are only lifted after users give permissions unrelated to the functionality (coded as \textit{forced access granting}). In related situations, we noticed \textit{forced dialogue interactions} that required users to engage with text elements otherwise occupying desired functionalities. More specific yet similar to the \textit{roach motel}~\cite{brignull_deceptive_nodate}, \textit{hard to cancel}~\cite{mathur2019}, or \textit{account deletion roadblocks}~\cite{gunawan_comparative_2021} dark patterns, a grace period of up to 30 days was announced by all SNSs before certain deletion processes were accepted. The \textit{Forced Grace Period} code was used in such instances. When deleting content, Facebook keeps the targeted item, as seen in Figure~\ref{fig:redirective-conditions-facebook}. Trying to bypass this rule, participants found it difficult to find the particular settings that would delete their content immediately. 
Interestingly, all four SNS denied users an immediate account deletion, as demonstrated by TikTok in Figure~\ref{fig:redirective-conditions-tiktok}. Each had a 30 days return option that would automatically reactivate accounts, even if users accidentally logged in to the SNS.

\section{Discussion \& Implications}
\begin{figure*}[h!]
    \centering
    \includegraphics[width=0.9\textwidth]{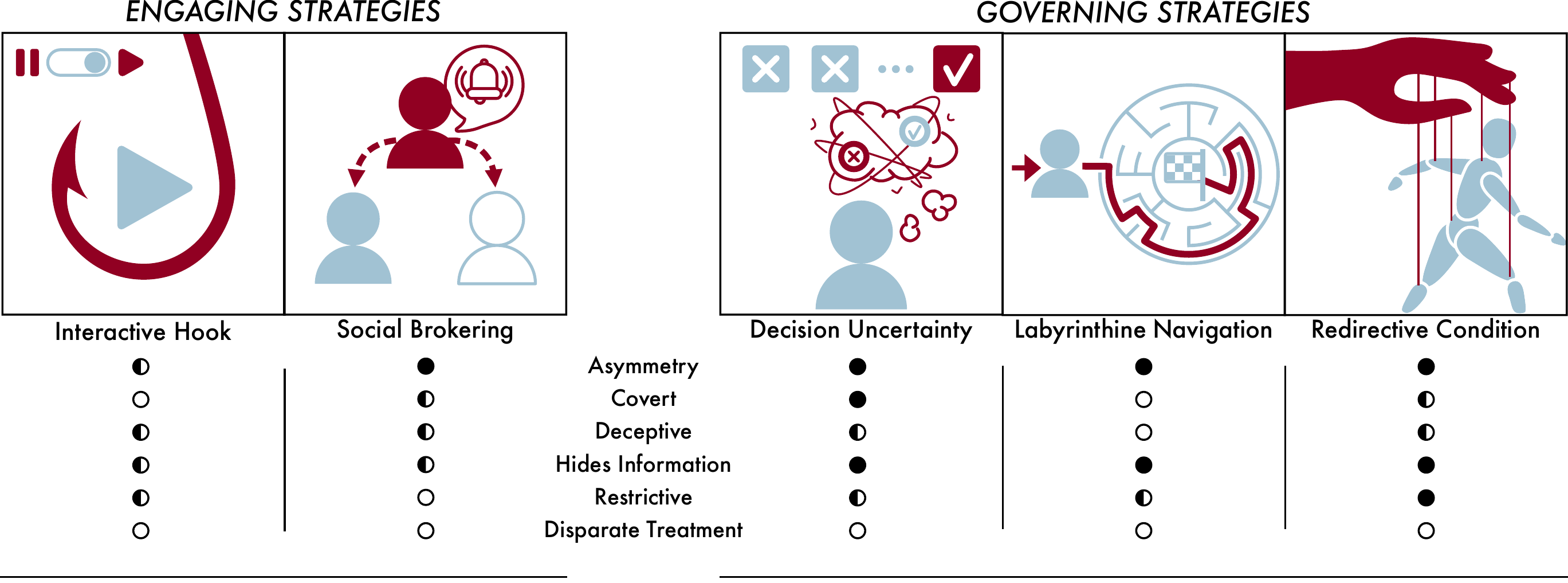}
    \caption{Summary of engaging and governing strategies with five SNS dark patterns in two strategies - engaging and governing. 
    For future work, each dark pattern was assigned corresponding attributes following Mathur et al.~\cite{mathur2019} six dark pattern characteristics.
    }
    \label{fig:pictograms-dark-pattern-strategies}
    \Description[Figure summarising the two strategies and five SNS-specific dark patterns.]{This figure shows contains five pictograms visually describing each of the SNS-specific dark patterns. Two, interactive hooks and social brokering, are placed under the engaging strategies. Three, decision uncertainty, labyrinthine navigation, and redirective conditions, are placed under the governing strategies.  The interactive hooks pictogram shows a red fishing hook surrounding a blue play button and a toggle that is set to a red auto-play option. The social brokering pictogram shows three people forming a triangle. Two are blue (one as fill-colour the other outlined to distinguish them), the third one is red and the upward peak of the triangle. From the red person, two arrows go towards each person establishing a connection. A speech bubble is over the third person's head, containing a ringing bell symbolising a notification. The decision uncertainty pictogram shows a sequence of choices in boxes at the top. Two boxes are blue containing an X, and the last box is red with a check symbol. Below is a blue person with a red thinking bubble above their head. Inside the bubble, confusion is symbolised by wavy lines and button options. These options are coloured in contrary colours to the ones on-top to indicate a mismatch in the person's mental model. The pictogram for labyrinthine navigation shows a circular labyrinth with a checkered flag in its centre. A blue person stands at the entry on the left-hand side. A red arrow indicates the direction into the labyrinth and is continued with a red line finding its way to the checkered flag. The redirective conditions pictogram shows a blue puppet. Each joint is connected with a red string that is connected to a larger red hand, indicating control over the puppet's movement. Each SNS-specific dark pattern is assigned Mathur et al.'s dark pattern characteristics (2021) by a full moon icon, if the characteristic applies, a half moon if it applies in certain situations but not always, and an empty circle if it does not apply. Characteristics for interactive hooks are: asymmetry = half moon; covert = circle; deceptive = half moon; hides information = half moon; restrictive = half moon; and disparate treatment = circle. Characteristics for social brokering are: asymmetry = full moon; covert = half moon; deceptive = half moon; hides information = half moon; restrictive = circle; and disparate treatment = circle. Characteristics for decision uncertainty are: asymmetry = full moon; covert = full moon; deceptive = half moon; hides information = full moon; restrictive = half moon; and disparate treatment = circle. Characteristics for labyrinthine navigation are: asymmetry = full moon; covert = circle; deceptive = circle; hides information = full moon; restrictive = half moon; and disparate treatment = circle. Characteristics for redirective condition are: asymmetry = full moon; covert = half moon; deceptive = half moon; hides information = full moon; restrictive = full moon; and disparate treatment = circle.}
\end{figure*}

Our examination of four SNSs --- Facebook, Instagram, TikTok, and Twitter --- identified instances of 44 dark patterns from a taxonomy of 80 codified in prior dark pattern literature. The work also identified instances of 22 inductive codes capturing malicious design artefacts that were not outlined in earlier research. Findings stem from expert reviews of the aforementioned SNSs carried out by six trained HCI researchers. Each researcher reviewed two SNSs and executed ten tasks designed to thoroughly understand the applications. The 22 inductive codes were then analysed, supported by axial coding. This resulted in the generation of five common themes describing SNS-specific dark patterns and two high-level strategies: Engaging strategies and governing strategies. Figure~\ref{fig:pictograms-dark-pattern-strategies} summarises these findings while further allocating Mathur et al.'s~\cite{Mathur2021} dark pattern characteristics.


\subsection{Dark Patterns In SNSs}
To explore the potential existence of dark patterns in SNSs, we began the thematic analysis by initialising a deductive dark pattern codebook comprising 80 dark pattern types from prior works~\cite{brignull_deceptive_nodate, conti_malicious_2010, zagal_dark_2013, greenberg_dark_nodate, bosch2016, gray2018, Gray2020a, mathur2019}. Answering our first research question regarding the types of dark patterns that are used across the four SNSs, we identified instances of 44 types. This might suggest that the other 36 types of dark patterns are, as our experience conducting thematic analysis attests, highly domain-specific, hindering their potential to be applied elsewhere. 

Indeed, despite the relative success of our attempts to apply these dark patterns to SNSs, different levels of generalisability and specification caused ambiguity decreasing confidence of coders when applying certain dark patterns. Some works provide alternating abstractions for their taxonomies. Conti and Sobiesk~\cite{conti_malicious_2010}, for instance, call their patterns malicious interface design \textit{techniques}, as the term dark pattern was not widely established at the time of their publication. In another work, Gray et al.~\cite{gray2018} speak of dark pattern \textit{strategies}, placing their findings on a more abstract level as, for example, Zagal et al.'~\cite{zagal_dark_2013} dark patterns, who speak of game dark patterns without further labels.

Because we agree with the necessity to understand dark patterns in different hierarchical contexts, we chose to include all taxonomies in our study to learn about differences when they are applied in situations outside their original field of application. Further guided by our decision to remain close to provided definitions, we learned that dark patterns with abstract definitions were applied more often throughout the SNSs. On the contrary, 36 dark patterns were left unused during our study. Certain dark patterns shared a name with varying definitions leading to ambiguity when used. Some dark patterns were derived from highly specific contexts, such as  e-commerce~\cite{brignull_deceptive_nodate, mathur2019} or games~\cite{zagal_dark_2013}, which hindered their usage elsewhere. Others contained overly precise descriptions descriptions~\cite{bosch2016} but could find usage elsewhere if phrased more generically. This implies that general and unspecific dark patterns gain the utility to describe unethical practices and, thus, better help to identify interface problems.

This implication is in line with prior research by Mathur et al.~\cite{mathur2019}. Aside from the identification of 12 dark patterns, the authors found an alternative approach to abstract the basic operations of dark patterns by mapping them onto cognitive biases that they exploit~\cite{mathur2019, Mathur2021}. The authors developed six characteristics distinguishing overall approaches to dark patterns: (1) asymmetric; (2) covert; (3) deceptive; (4) hides information; (5) restrictive; and (6) disparate treatment. Through these characteristics, the authors were able to characterise a comprehensive dark pattern taxonomy in which we notice an accessible and extendable framework for future works.

\subsection{SNS-Specific Dark Patterns \& Strategies}

Answering the second research question, asking about dark patterns unique to SNSs, our study revealed five types of dark patterns not contained in previous work. When defining the dark patterns, one goal was to provide enough abstraction to enable applicability outside SNSs. For the same reason, we placed them in Mathur et al.'s~\cite{mathur2019, Mathur2021} six characteristics to allow future works to consider these dark patterns under their lens as well. As can be seen in Figure~\ref{fig:pictograms-dark-pattern-strategies}, not all of Mathur et al.'s characteristics could be applied with the same effectiveness or explicitness. Interestingly, we were not able to apply the \textit{disparate treatment} characteristic at all because it solely contains Zagal et al.'s~\cite{zagal_dark_2013} game dark patterns which describe mechanics where games benefit players who, for example, buy advantages. We found none of the four considered SNSs to offer distinct benefits to certain users while posing disadvantages to others. 

Another reason why Mathur et al.'s characteristics show lower effectiveness in SNSs may be described in how the two strategies operate, distinguishing them from prior dark pattern settings. As an incentive to increase time spent on their platforms, SNS need satisfied users. Since the implementation of harmful designs could act as an antagonist to this goal, jeopardising potential advertisement revenue, we describe engaging and governing strategies that navigate users' decision-making while possibly keeping their satisfaction with the SNS high~\cite{mildner_ethical_2021}. Although we cannot draw an easy causal connection between how dark patterns affect users and users' well-being, assessing identified SNS dark patterns further resulted in the description of two design strategies, \textit{engaging strategies} and \textit{governing strategies} (see Figure \ref{fig:pictograms-dark-pattern-strategies}). While one group aims to increase the time users spend on SNS, the other governs their decision-making by nudging users into desired directions or keeping them from other options by, for example, designing for increased friction.

With the intent to enable future work to better build on our findings and inspired by the types of dark patterns that showed more pertinence in our study, our aim was to describe our dark patterns to allow application outside SNS-specific contexts. As a different environment, many games feature achievement tracks requiring players to regularly play the game to progress. Similar to Zagal et al.'s \textit{playing by appointment}, it is imaginable that these games feature \textit{interactive hooks} to keep players engaged or contain distracting elements obfuscating users' decisions and, thus, deploying \textit{decision uncertainty}.

Although this work mainly contributes to the academic community, this strand of research has gained traction outside, including regulation and guidelines.
In a first line of defence against violations of GDPR~\cite{_gdpr_2016} requirements in SNS contexts, the EDPB~\cite{edpb_guidelines_2022} developed a guideline for designers and users to recognise and avoid dark patterns. However, the included dark pattern categories (overloading, skipping, stirring, hindering, fickle, and left in the dark) lack alignment with taxonomies developed in the academic community that relies on empirical evidence. In this regard, we acknowledge a potential for better cooperation between law and HCI research to work on the protection of users in a joint effort. To offer some aid in these efforts, our goal was to create definitions that are broadly applicable. The five dark patterns expand current taxonomies by the scope of SNSs, whereas the two strategies provide high-level categories to describe alternate burdens placed on users not yet covered. 




\section{Limitations \& Future Work}
As previously stated in Section~\ref{sec:expert_review}, we faced certain limitations in the data collection phase of this work. Firstly, we decided to focus on only four SNSs that we deemed comparable after prior considerations looking at a wider range of applications for the user count and similar functionalities. This limitation extends to the fact that we limited the review to only looking at their mobile applications. Although this limitation allowed us to study each SNS in a single modality on a deeper level, our findings do not necessarily represent all SNSs. None of the considered SNSs indicates any usage of the \textit{disparate treatment} characteristic, which describes discrepancies in the treatment of paid and free customers, respectively. This might be more relevant in a study of LinkedIn, for example, as it offers both free and paid subscriptions, the latter of which affords certain advantages like hiding one's identity when looking at another user's profile. Future work could extend our findings by considering other SNSs. Secondly, we decided early on to only focus on mobile applications of considered SNSs, following users' preferences. However, users may be faced with alternative dark patterns based on SNSs' modalities, as the research by Guanawa et al.~\cite{gunawan_comparative_2021} or Schaffner et al.~\cite{schaffner_understanding_2022} suggest. Future studies could consider our results when expanding on other SNSs and modalities. Thirdly, this research was conducted during the COVID-19 pandemic. The data collection was thus conducted without supervision after providing necessary information, including a comprehensive manual. While neither screen nor voice recordings indicated confusion about the tasks, room for misunderstanding remained. Thirdly, we relied on HCI researchers as expert reviewers. This decision is justified by their competence to understand and recognise state-of-the-art design practices, unlike regular users would be able to. However, by limiting reviewers' expertise to a single domain, we may have missed alternative expertise that may surface additional findings. Future work could consider recruiting experts with backgrounds in cognitive science and psychology, including a thorough understanding of cognitive biases. Their expertise could be particularly interesting in establishing connections between our current understanding of dark patterns and of cognitive biases.

During the thematic analysis, we also faced some limitations. Firstly, we limited the deductive codebook to eight works comprising 81 dark patterns. We relied on Mathur~\cite{Mathur2021} dark pattern review to get a comprehensive taxonomy. However, we decided to neglect dark patterns that were described outside the academic community, including guidelines such as those published by the NCC~\cite{ncc_2018}, CNIL~\cite{cnil_2019}, or the EDPB~\cite{edpb_guidelines_2022}. Moreover, dark patterns by Gunawan~\cite{gunawan_comparative_2021} were not included, as they were not mapped onto Mathur et al.~\cite{mathur2019, Mathur2021} dark pattern characteristics, making comparisons more difficult. Future work could generate a complete mapping of a comprehensive dark pattern taxonomy onto Mathur et al.'s characteristics. Connecting dark patterns to cognitive biases, as Mathur et al. propose, seems to be a natural next step for the dark pattern discourse. To prepare this work for this direction, we assigned our findings to appropriate characteristics. In this process, we noticed that the current scope of characteristics does not cover all problematic design strategies contained in SNS. However, a thorough analysis of potential cognitive biases active in SNS would have been outside the scope of this research. Future work could address this gap and extend this work by bringing together the studies surrounding cognitive biases and dark patterns to generate a resourceful and sustainable framework.


\section{Conclusion}
In recent years, the dark pattern landscape has expanded into various different domains. In this paper, we contribute to this body of research through the application and expansion of a comprehensive taxonomy of dark patterns in the context of SNSs. Supported by thematic analysis, we investigate Facebook, Instagram, TikTok, and Twitter and confirm that the platforms all deployed a variety of dark patterns and design strategies aimed at limiting users' agency, steering their decision-making. Findings suggest that dark patterns with higher grades of abstraction are easier to be applied in multiple contexts compared to those given narrower definitions. Lastly, our results yield evidence for two high-level strategies - engaging and governing - containing five types of dark patterns previously not described.

\begin{acks}
The research of this work was partially supported by the Klaus Tschira Stiftung gGmbH.
\end{acks}

\bibliographystyle{ACM-Reference-Format}
\bibliography{references.bib}
\newpage

\appendix
\section{Inductive Codebook Including Descriptions}\label{app:inductive-codebook}

\begin{table*}[!htpb]
\centering
\begin{tabular}{p{.02\textwidth}p{.26\textwidth}p{.6\textwidth}}
\toprule
    & \textbf{Code}                 & \textbf{Description} \\ \hline 
\rowcolor[HTML]{EFEFEF} 
1   & Addictive Design              & Features or elements that keep users hooked to content.\\
2   & Auto Accept Third Party Terms & Unknowingly giving consent to share data with third parties per default settings.\\\rowcolor[HTML]{EFEFEF} 
3   & Autoplay Content              & Auto-playing content without further actions by the user.\\
4   & Clinging To Accounts          & Making the process of deleting an account unnecessarily difficult or reactivating accounts after deletion has already been initialised.\\\rowcolor[HTML]{EFEFEF} 
5   & Decision Governing            & Interface instances that navigate or steer users' decision-making.\\
6   & Decision Uncertainty          & Users do not know what they are left in confusion and what consequences their decision will have.\\\rowcolor[HTML]{EFEFEF} 
7   & External Solution Search      & Not able to find specific features or settings, users fall back to use search engines to find what they are looking for.\\
8   & False Content Customisation   & Shown content does not fit the users' preferences or followed accounts.\\\rowcolor[HTML]{EFEFEF} 
9   & Fear Of Missing Out           & Feeling pressured to (re)visit specific SNS features out of fear to miss something.\\
10  & Forced Access Granting        & Features requiring unnecessary access to special device hardware or local data.\\\rowcolor[HTML]{EFEFEF} 
11  & Forced Dialogue Interaction   & Prompting Text Boxes that require immediate attention with no option to dismiss them without interaction.\\
12  & Forced Grace Period           & When deleting content or accounts, users are forced to wait a minimum amount of days before changes become active (often 30 days).\\\rowcolor[HTML]{EFEFEF} 
13  & Gamification                  & Playful elements that motivate users to do something by presenting artificial progress and/or rewards to get more data.\\
14  & Hidden In Plain Sight         & Crucial information is often obscured by attention-grabbing interface elements.\\\rowcolor[HTML]{EFEFEF} 
15  & Infinite Scrolling            & Users can infinitely scroll through content (often, more and more suggested content is shown the more they progress.\\
16  & Labyrinth                     & Nested interface structures users get easily lost in.\\\rowcolor[HTML]{EFEFEF} 
17  & Persuasive Language           & Emotional pressuring language to push towards a specific direction that may not be in the users best interest. \\
18  & Plain Evil                    & Interface elements suddenly change functionalities or are being exchanged for alternative ones.\\\rowcolor[HTML]{EFEFEF} 
19  & Pull To Refresh               & Pulling downward on a content-displaying interfaces will load new content. Sometimes, suggested content is shuffled into the feed to give users more to look at.\\
20  & Reduced Friction              & Purposefully making certain elements easier accessible and thus pushing alternatives into the back.\\\rowcolor[HTML]{EFEFEF}
21  & Regression Toward The Mean    & Users are suggested topics which most people like and thus are likely to add to the already existing ``popular main topics''.\\
22  & Social Connector              & Requesting additional information to connect to friends/family and other social circles.\\\rowcolor[HTML]{EFEFEF} 
\bottomrule
\end{tabular}
\caption{Table containing all 22 codes from the inductive codebook from study 1, later used to create five themes deriving into the SNS-specific dark patterns.}
\end{table*}

\end{document}